\documentclass[fleqn,usenatbib]{mnras}
\usepackage[T1]{fontenc}
\DeclareRobustCommand{\VAN}[3]{#2}
\let\VANthebibliography\thebibliography
\def\thebibliography{\DeclareRobustCommand{\VAN}[3]{##3}\VANthebibliography}
\usepackage{float}

\usepackage{graphicx}	
\usepackage{amsmath}	
\usepackage{amssymb}	
\usepackage{newtxtext,newtxmath}

\title[The ASAS-SN Catalog of Variable Stars X]{The ASAS-SN Catalog of Variable Stars X: Discovery of 116,000 New Variable Stars Using g-band Photometry}

\author[C. T. Christy et al.]{C. T. Christy,$^{1}$\thanks{E-mail: christy.125@osu.edu}
T. Jayasinghe$^{1,2}$,
K. Z.  Stanek$^{1,2}$,
C. S. Kochanek$^{1,2}$,
T. A. Thompson$^{1,2}$,
\newauthor
B. J. Shappee$^{3}$,
T.~W.-S.~Holoien$^{4,\thanks{NHFP Einstein Fellow}}$,
J. L. Prieto$^{5}$,
Subo Dong$^{6}$,
W. Giles$^{7}$\\
$^{1}$ Department of Astronomy, The Ohio State University, 140 West 18th Avenue, Columbus, OH 43210, USA\\
$^{2}$ Center for Cosmology and Astroparticle Physics, The Ohio State University, 191 W. Woodruff Avenue, Columbus, OH 43210 \\
$^{3}$ Institute for Astronomy, University of Hawai’i, 2680 Woodlawn Drive, Honolulu, HI 96822, USA \\
$^{4}$ The Observatories of the Carnegie Institution for Science, 813 Santa Barbara St., Pasadena, CA 91101, USA \\
$^{5}$ Núcleo de Astronomía de la Facultad de Ingeniería y Ciencias, Universidad Diego Portales, Av. Ejército 441, Santiago, Chile \\
$^{6}$ Kavli Institute for Astronomy and Astrophysics, Peking University, Yi He Yuan Road 5, Hai Dian District, China\\
$^{7}$ ASC Technology Services, 433 Mendenhall Laboratory 125 South Oval Mall Columbus OH, 43210, USA\\
}

\date{Accepted XXX. Received YYY; in original form ZZZ}

\pubyear{2022}

\begin{document}
\label{firstpage}
\pagerange{\pageref{firstpage}--\pageref{lastpage}}
\maketitle

\begin{abstract}
\indent The All-Sky Automated Survey for Supernovae (ASAS-SN) is the first optical survey to monitor the entire sky, currently with a cadence of $\lesssim 24$ hours down to $g \lesssim 18.5$ mag. ASAS-SN has routinely operated since 2013, collecting $\sim$ 2,000 to over 7,500 epochs of $V$ and $g-$band observations per field to date. This work illustrates the first analysis of ASAS-SN's newer, deeper, higher cadence $g-$band data. From an input source list of ${\sim}55$ million isolated sources with $g<18$~mag, we identified $1.5\times10^6$ variable star candidates using a random forest classifier trained on features derived from \textit{Gaia}, 2MASS and AllWISE. Using ASAS-SN $g-$band light curves, and an updated random forest classifier augmented with data from Citizen ASAS-SN, we classified the candidate variables into 8 broad variability types. We present a catalog of ${\sim}116,000$ new variable stars with high classification probabilities, including  ${\sim}111,000$ periodic variables and ${\sim}5,000$ irregular variables. We also recovered ${\sim}263,000$ known variable stars. 
\end{abstract}


\section{Introduction}

Variable stars are an important and dynamic area of modern astronomical research. Variability provides extra observational information (periods, amplitudes, etc.) which can be used to determine physical parameters such as mass, radius, luminosity, and rotation rates (e.g. \citealt{Percy}). For example, $\delta$ Scuti variables allow us to study the scaling relations between stellar parameters (effective temperature, surface gravity, density, etc.) and astroseismology (e.g. \citealt{dsct}). Eclipsing binaries allow us to accurately measure stellar parameters such as mass and radius (e.g. \citealt{Torres_2009}). RR Lyrae variables have been used to derive the structural parameters of the inner halo and thick disc of the Milky Way (e.g. \citealt{2018MNRAS.479..211M}). The period luminosity-relationship of Cepheid variables allow distance estimates on cosmic scales (e.g. \citealt{1908AnHar..60...87L}). In short, variable stars are used to solve astrophysical problems, especially those requiring knowledge of distances, stellar structure, and populations (e.g. \citealt{2014IAUS..298...40F}). 

Continuity of coverage and continuing surveys are particularly important for finding rare systems that can become ``Rosetta stones'' for stellar processes. In the modern era, large surveys such as the All-Sky Automated Survey (ASAS; \citealt{pojmanski_2002}), the All-Sky Automated Survey for SuperNovae (ASAS-SN;  \citealt{2014ApJ...788...48S,2017PASP..129j4502K,Jayasinghe2018,Jayasinghe2021}), the Asteroid Terrestrial-impact Last Alert System (ATLAS; \citealt{Heinze_2018}; \citealt{Tonry_2018}), the Catalina Real-Time Transient Survey (CRTS; \citealt{Drake_2009}), EROS (\citealt{Derue_2002}), \textit{Gaia} \citep{2016, 2018}, MACHO \citealt{Alcock_2000},  the Northern Sky Variability Survey (NSVS; \citealt{Woniak_2004}), the Optical Gravitational Lensing Experiment (OGLE; \citealt{udalski_2004}), and the Zwicky Transient Facility (ZTF; \citealt{bellm2014zwicky}) have rapidly advanced the collection of known variables to over $\sim2.1 \times 10^6$ examples based on the American Association of Variable Star Observers (AAVSO) catalog \citep{2006SASS...25...47W}. 

ASAS-SN was originally designed to study bright supernovae and other transients such as tidal disruption events, cataclysmic variables, AGN, and stellar flares \citep{Holoien_2016}. ASAS-SN data are also well-suited for the cataloging, classification, and study of variable stars. For the initial \textit{V}-band catalog of variables, the light curves for ${\sim} 60$ million stars were classified through machine learning techniques, resulting in a catalog of ${\sim}426,000$ variables, of which ${\sim} 220,000$ were new discoveries \citep{2020arXiv200610057J,Jayasinghe2021}. We are now using citizen science through the Citizen ASAS-SN project hosted on the Zooniverse\footnote{Zooniverse:https://www.zooniverse.org/} \citep{CitizenRN} to identify and classify variables in the $g-$band data. The ASAS-SN citizen science campaign has also already begun to identify a host of new variable stars in our data \citep{CitizenDR1}.

In Paper I \citep{Jayasinghe2018}, we discovered ${\sim}66,000$ new variables that were flagged during the search for supernovae and homogeneously analyzed ${\sim} 412,000$ known variables from the VSX catalog in Paper II \citep{Jayasinghe2019a}. In Paper III \citep{Jayasinghe2019b}, we characterized the variability of ${\sim}1.3$ million sources in the southern TESS (Transiting Exoplanet Survey Satellite; \citealt{2015JATIS...1a4003R}) continuous viewing zone and identified ${\sim} 11,700$ variables, including ${\sim} 7,000$ new discoveries. In Paper IV \citep{2019MNRAS.487.5932P}, we have also explored the synergy between ASAS-SN and large scale spectroscopic surveys using data from APOGEE \citep{2015AJ....150..148H} with the discovery of the first likely non-interacting binary composed of a black hole with a field red giant \citep{2019Sci...366..637T}. In Paper V, we identified ${\sim}220,000$ variable sources with $V<17$ mag in the southern hemisphere, of which ${\sim}88,300$ were new discoveries \citep{Jayasinghe2019c}. In Paper VI, we derived period--luminosity relationships for $\delta$ Scuti stars \citep{Jayasinghe2020b}. We studied contact binaries in Paper VII \citep{Jayasinghe2020a}. In Paper VIII, we identified 11 new ``dipper'' stars in the Lupus star forming region \citep{2020MNRAS.496.3257B}. In Paper IX, we used spectroscopic information from LAMOST, GALAH, RAVE, and APOGEE to study the physical and chemical properties of these variables \citep{2021MNRAS.503..200J}.

In this paper, we present the first all-sky catalog of variables detected in the newer, deeper, higher cadence $g-$band ASAS-SN data. The complete list of the crossmatched variables and the ASAS-SN discoveries along with their $g-$band light curves are provided online at the ASAS-SN Variable Stars Database (\url{https://asas-sn.osu.edu/variables}) and have been reported to the AAVSO. Section $\S 2$ details the data and methods used to identify and classify the variable star candidates. In Section $\S 3$ we discuss the results and in Section $\S 4$ we present our conclusions.

\section{Observations and Methods}
\subsection{Data}
In 2014, ASAS-SN began surveying the entire sky in the $V-$band with a limiting magnitude of $V \lesssim 17$ mag and a $\sim 2-3$ day cadence using 8 telescopes on two mounts in Chile and Hawaii \citep{2014ApJ...788...48S,2017PASP..129j4502K}. Since 2018, ASAS-SN has shifted to using a $g$-band filter and expanded to 20 cameras on 5 mounts, adding new units in South Africa, Texas, and Chile \citep{Jayasinghe2018}. All of the ASAS-SN telescopes are hosted by the Las Cumbres Observatory (LCO; \citealt{Brown_2013}). When compared to the $V-$band data, the $g$-band data have an improved depth ($g \lesssim 18.5$ mag), cadence ($\lesssim 24$ hours in the $g$-band vs. ${\sim}2-3$ days in the $V$-band), and reduced diurnal aliasing due to the longitudinal spread of the ASAS-SN units. The ASAS-SN $V-$band observations were made by the ``Brutus'' (Haleakala, Hawaii) and ``Cassius'' (CTIO, Chile) quadruple telescopes between 2013 and 2018. Our $g-$band observations added data from the ``Payne'' (Sutherland, South Africa), ``Bohdan'' (CTIO, Chile), and ``Leavitt'' (McDonald, Texas) quadruple telescopes starting in 2018. Each ASAS-SN camera takes 3 images with 90 second exposures for each epoch. The field of view of an ASAS-SN camera is 4.5 deg$^2$, the pixel scale is 8\farcs0, and the FWHM is typically ${\sim}2$ pixels. ASAS-SN saturates at $g{\sim} 11-12$ mag, but we attempt to correct the light curves of saturated sources for the bleed trails (see \citealt{2017PASP..129j4502K}). The $g-$band light curves were extracted as described in \citet{Jayasinghe2018} using image subtraction \citep{1998ApJ...503..325A,2000A&AS..144..363A} and aperture photometry on the subtracted images with a 2 pixel radius aperture. We corrected the zero point offsets between the different cameras as described in \citet{Jayasinghe2018}. The photometric errors were recalculated as described in \citet{Jayasinghe2019b}.

\subsection{Identifying Variable Star Candidates}

We started with the \verb"refcat2" catalog \citep{2018ApJ...867..105T} as our input source catalog. We selected ${\sim}54.8$ million \verb"refcat2" sources with $g<18$ mag and \verb"r1"$>30"$, where \verb"r1" is the radius at which the cumulative $G$ flux in the aperture exceeds the flux of the source being considered and is a measure of the crowding around a star. We use the limit on \verb"r1" to reduce the number of heavily blended sources. 

The production of 55 million of light curves is computationally expensive, so we used external photometry from \textit{Gaia} EDR3, 2MASS and AllWISE to identify likely variable sources rather than simply generating light curves for every source. In particular, the photometric uncertainties in \textit{Gaia} EDR3 encodes information about the photometric variability of sources (see, for e.g., \citealt{Andrew2021}). At a fixed \textit{G} magnitude, variable stars in \textit{Gaia} have larger photometric uncertainties than constant stars and therefore, we can use the photometric uncertainties available in \textit{Gaia} EDR3 as a feature to identify stellar variability.

We built a variability classifier based on a random forest (RF) model with \verb"scikit-learn"  \citep{2012arXiv1201.0490P}. The goal was to first divide the stars into two groups: CONST (constant stars) and VAR (potential variables). The variable star candidates will then be analyzed in detail, so it is more important not to lose real variables than to accidentally include non-variables. For the training set, we used the ${\sim}204,000$ known variables used to train the RFC variability classifier in \citet{CitizenDR1} and the variables identified in Citizen ASAS-SN DR1. For the constant sources, we used a set of ${\sim}250,000$ non-variable sources identified in Citizen ASAS-SN DR1. We used 16 features from \textit{Gaia} EDR3, 2MASS and AllWISE. These include the EDR3 \textit{G}, \textit{BP}, \textit{RP} magnitudes and the associated uncertainties, the $BP-RP$ color, the $BP-RP$ excess factor, signal-to-noise ratios in \textit{G} and \textit{BP}, the renormalized unit weight error (RUWE), the $J-K_s$ color, the absolute $W_{RP}$ magnitude and the absolute $W_{JK}$ magnitude. The EDR3 signal-to-noise ratios are essentially the ratio of the observed flux divided by the error in the flux. As noted earlier, the EDR3 photometric uncertainties and flux errors encode information about the photometric variability of stars. We also used the absolute, ``reddening-free'' Wesenheit magnitudes \citep{Madore1982,Lebzelter2018} \begin{equation}
    W_{RP}=M_{\rm RP}-1.3(BP-RP) \,, 
	\label{eq:wrp}
\end{equation} 
and
\begin{equation}
    W_{JK}=M_{\rm K_s}-0.686(J-K_s) \,
	\label{eq:wk}
\end{equation} and the probabilistic EDR3 distances from \citet{Bailer-Jones2021}.

\begin{figure}
    \centering
    \includegraphics[width =  0.5\textwidth]{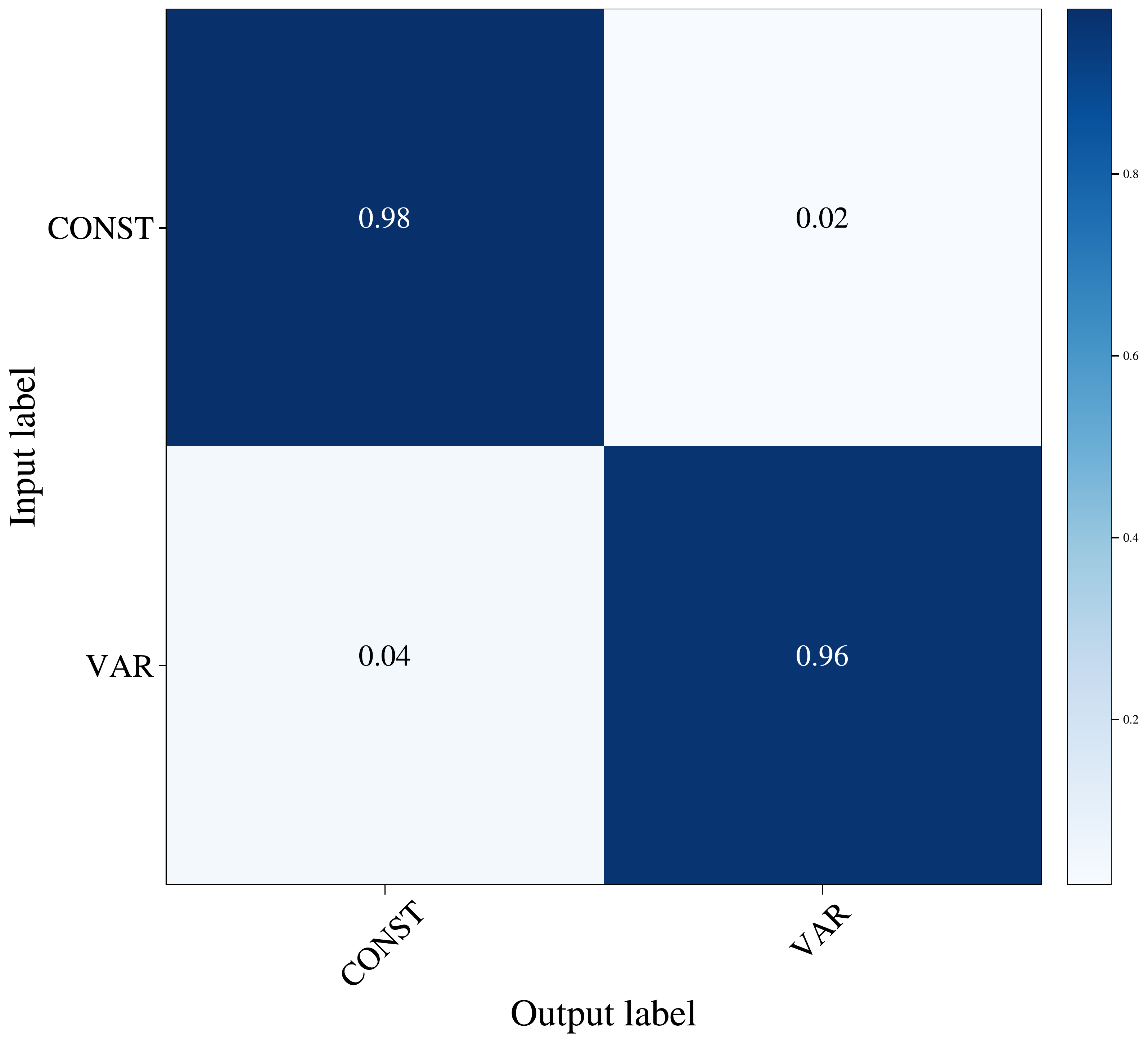}
    \caption{Confusion matrix for the VAR/CONST classifier trained on features from \textit{Gaia} EDR3, 2MASS and AllWISE.}
    \label{fig:cfm_varconst}
\end{figure}

The parameters of the random forest model were optimized using cross-validation to maximize the overall $F_1$ score of the classifier. The number of decision trees in the forest was initialized to \verb"n_estimators="1200. We also limited the maximum depth of the decision trees to \verb"max_depth=16" in order to mitigate over-fitting, set the number of samples needed to split a node as \verb"min_samples_split=10" and set the number of samples at a leaf node as \verb"min_samples_leaf=5". To further minimize over-fitting, we also assigned weights to each class with \verb"class_weight=`balanced_subsample'". For any given source, the RF classifier assigns classification probabilities $\rm Prob(Const)$ and $\rm Prob(Var)=1-Prob(Const)$. The output classification of the RF classifier is the class with the highest probability. We split the training sample, using $90\%$ for training and $10\%$ for testing, in order to evaluate the performance of the RF classifier. The confusion matrix for the trained RF model is shown in Figure \ref{fig:cfm_varconst}. The greatest confusion (4\%) arises from input variable sources that are subsequently classified as constant stars. The overall precision, recall and $F_1$ parameters for the classifier are 97.3$\%$, 97.1$\%$ and 97.2$\%$ respectively. We then classified the  ${\sim}54.8$ million \verb"refcat2" sources using the trained RF classifier, and we identified ${\sim}1.48$ million variable star candidates (${\sim}2.7\%$). We extracted ASAS-SN $g-$band light curves of these candidates and determined periods using only the GLS periodogram as in Citizen ASAS-SN (see \citealt{CitizenDR1}).

\subsection{Classifying Variable Stars}

\begin{figure}
    \centering
    \includegraphics[width =  0.5\textwidth]{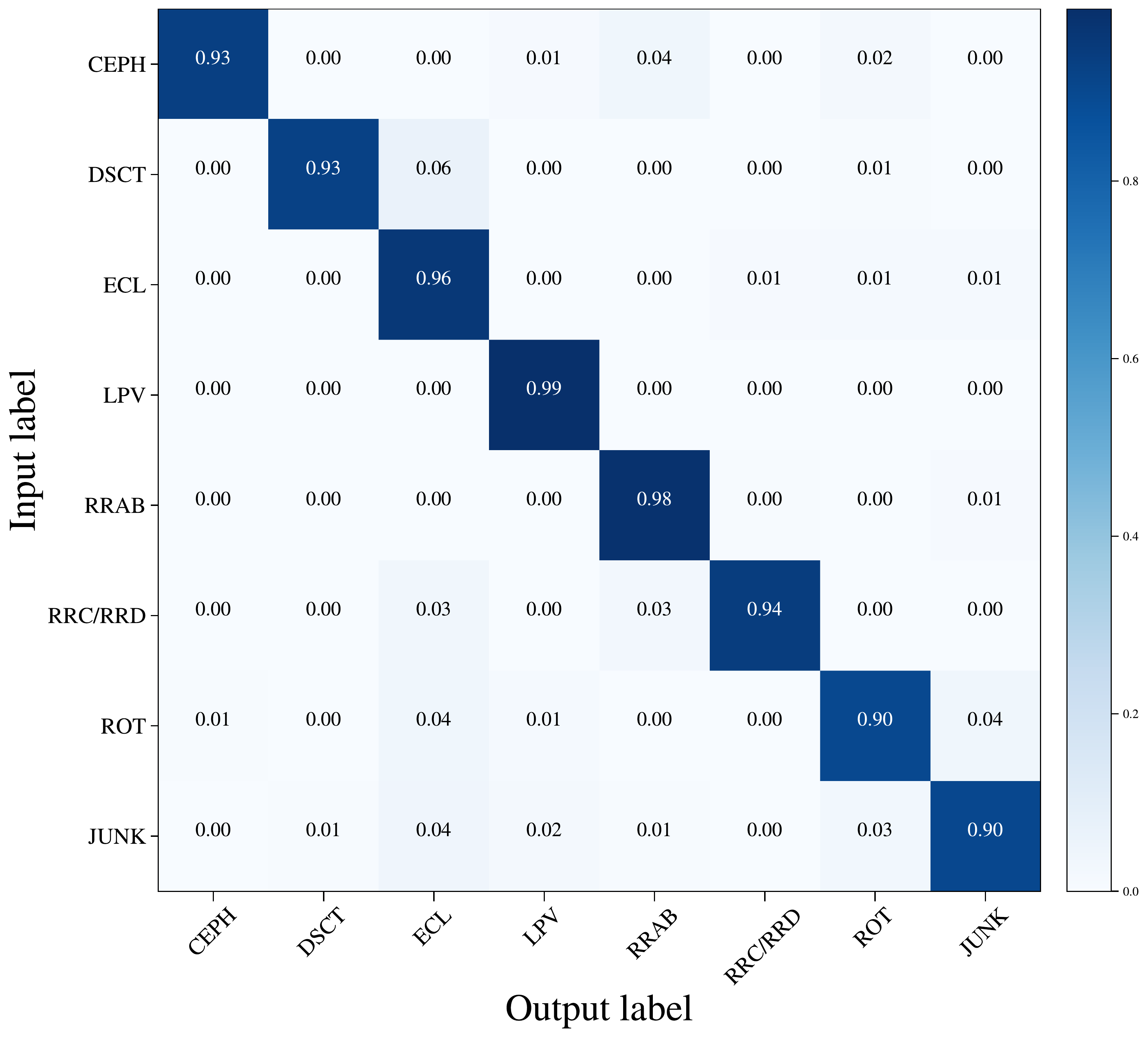}
    \caption{Confusion matrix for the updated classifier from \citet{CitizenDR1}. In the updated version used in this work, we included the ``JUNK'' class to identify light curves with spurious variability.}
    \label{fig:cfmrfcvars}
\end{figure}

We retrained the $g$-band random forest classifier described in \citet{CitizenDR1} to include a new category for spurious ``JUNK'' variables. The training set for the JUNK class was the ${\sim}12,000$ ``JUNK'' variables identified by citizen scientists in the first data release from Citizen ASAS-SN \citep{CitizenDR1}. The updated RF classifier classifies sources into 8 broad classes (CEPH, DSCT, ECL, LPV, RRAB, RRc/RRd, ROT and JUNK) which are subsequently refined into sub-classes (see \citealt{Jayasinghe2019a}). The overall precision, recall and $F_1$ parameters for the updated RF classifier are 91.6$\%$, 94.2$\%$ and 92.8$\%$ respectively. The confusion matrix for the trained RF model is shown in Figure \ref{fig:cfmrfcvars}.

\begin{figure}
    \centering
    \includegraphics[width = 0.48\textwidth]{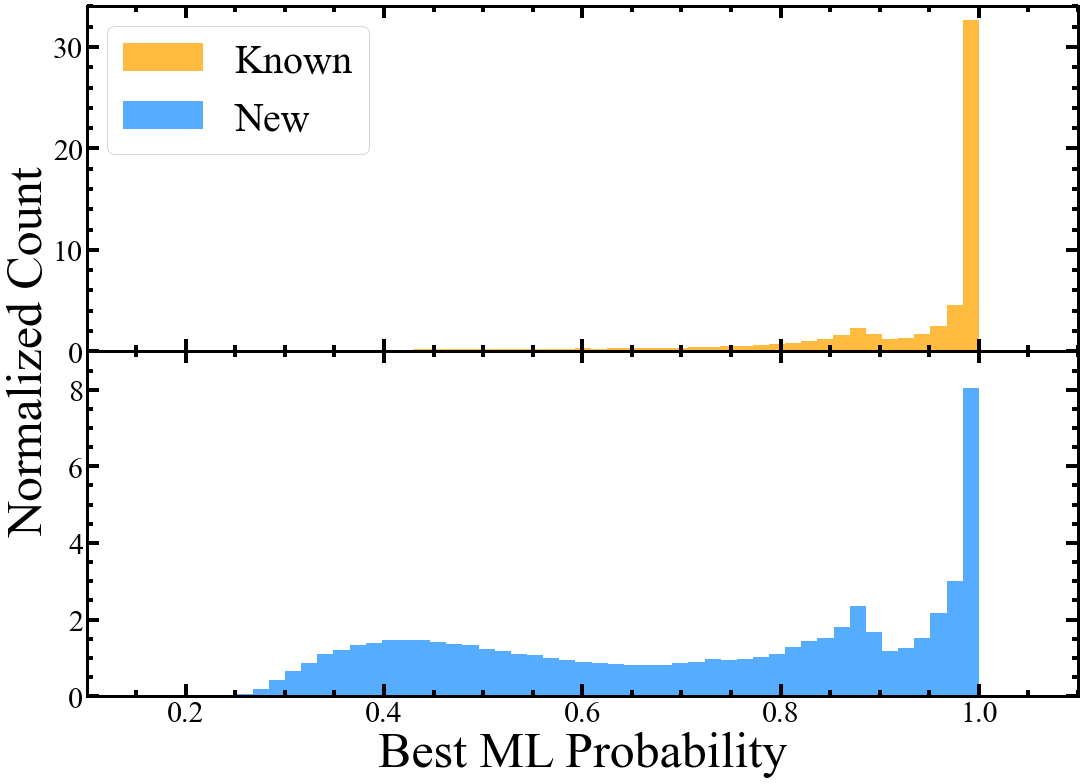}
    \caption{Machine learning probability distribution for the full input list of ${\sim}755,000$ variables separated into known and new sources.}
    \label{fig:ML_probs}
\end{figure}

We applied the updated variability classifier to the ${\sim}1.48$ million variable star candidates. To reduce the number of false positives, we imposed a probability cut of $\rm Prob>0.95$ for variables with periods close to diurnal and lunar aliases (e.g., 1 d, 2 d, 29 d, 60 d). Additionally, to eliminate false positives caused by spurious variability, we used a probability cut of $\rm Prob>0.8$ for short period DSCT variables. Following this step, we are left with ${\sim}755,000$ variables.

We cross-matched the list of ${\sim}755,000$ variables with the AAVSO VSX \citep{2006SASS...25...47W}, OGLE III \citep{poleski2012optical}, and OGLE IV \citep{kozlowski2013supernovae,Soszynski2014,Udalski2015,Soszynski2015,Soszynski2016,Udalski2018,Pietrukowicz2020,Soszynski2021} catalogs using a matching radius of 16 arcsec and found 359,265 previously known variables. The VSX catalog contains all the variables previously identified by many wide field surveys including ASAS-SN \citep{2020arXiv200610057J}, ATLAS \citep{Heinze2018}, WISE \citep{Chen2018}, and ZTF \citep{Chen2020}. After excluding the known variables, there are 395,494 new variables in our list. 

\begin{table*}
\centering
\begin{tabular}{llrrr}
\hline
RF Classification & Description   & $\rm N_{\text{Known}}$ & $\rm N_{\text{New}} $ & $\rm N_{\text{New}}/\rm N_{\text{Known}}$ \\ \hline
CWA 	 & W Virginis  type variables with P $>$ 8 d 	 	         & 153 	 & - 	 & - \\
CWB 	 & W Virginis  type variables with P $<$ 8 d 	 	         & 73 	 & 1 	 & 0.01 \\
DCEP 	 & $\delta$ Cephei-type classical Cepheid variables 	 	 & 432 	 & 2 	 & <0.01 \\
DCEPS 	 & First overtone Cepheid variables 	 	                 & 109 	 & 3 	 & 0.03 \\
DSCT 	 & $\delta$ Scuti type variables 	 	 & 848 	 & 1547 	 & 1.82 \\
EA 	 & Detached Algol-type binaries 	 	 & 17447 	 & 4480 	 & 0.26 \\
EB 	 & $\beta$ Lyrae-type binaries 	 	 & 11820 	 & 1551 	 & 0.13 \\
EW 	 & W Ursae Majoris type binaries 	 	 & 42737 	 & 4833 	 & 0.11 \\
GCAS 	 & $\gamma$ Cassiopeiae variables 	 	 & - 	 & - 	 & - \\
HADS 	 & High amplitude $\delta$ Scuti type variables 	 	 & 1725 	 & 506 	 & 0.29 \\
L 	 & Irregular Variables 	 	 & 9152 	 & 4786 	 & 0.52 \\
M 	 & Mira variables 	 	 & 6287 	 & 363 	 & 0.06 \\
ROT 	 & Spotted Variables with rotational modulation 	 	 & 14755 	 & 38414 	 & 2.60 \\
RRAB 	 & Fundamental Mode RR Lyrae variables 	 	 & 18455 	 & 580 	 & 0.03 \\
RRC 	 & First Overtone RR Lyrae variables 	 	 & 6518 	 & 450 	 & 0.07 \\
RRD 	 & Double Mode RR Lyrae variables 	 	 & 482 	 & 30 	 & 0.06 \\
RVA 	 & RV Tauri variables (Subtype A) 	 	 & 3 	 & - 	 & - \\
SR 	 & Semi-regular variables 	 	 & 131479 	 & 57925 	 & 0.44 \\
YSO 	 & Young Stellar Objects 	 	 & 209 	 & 147 	 & 0.70 \\
VAR 	 & Variable star of unspecified type 	 	 & 150 	 & 409 	 & 2.73 \\ \hline
\textbf{Total} & & 262834 & 116027 & 0.44 \\\hline                               
\end{tabular}
\caption{ML Classification breakdown of the variables from this search.}
\label{tab:1}
\end{table*}

\subsection{Quality Control}
In Figure \ref{fig:ML_probs}, we show the distribution of the highest probability assigned to each variable candidate's classification for the known variables and the new candidates. The probability distribution for the known variables is more concentrated towards unity than the new candidates. The differences indicated a need for additional sample restrictions.

\begin{figure}
    \centering
    \includegraphics[width = 0.48\textwidth]{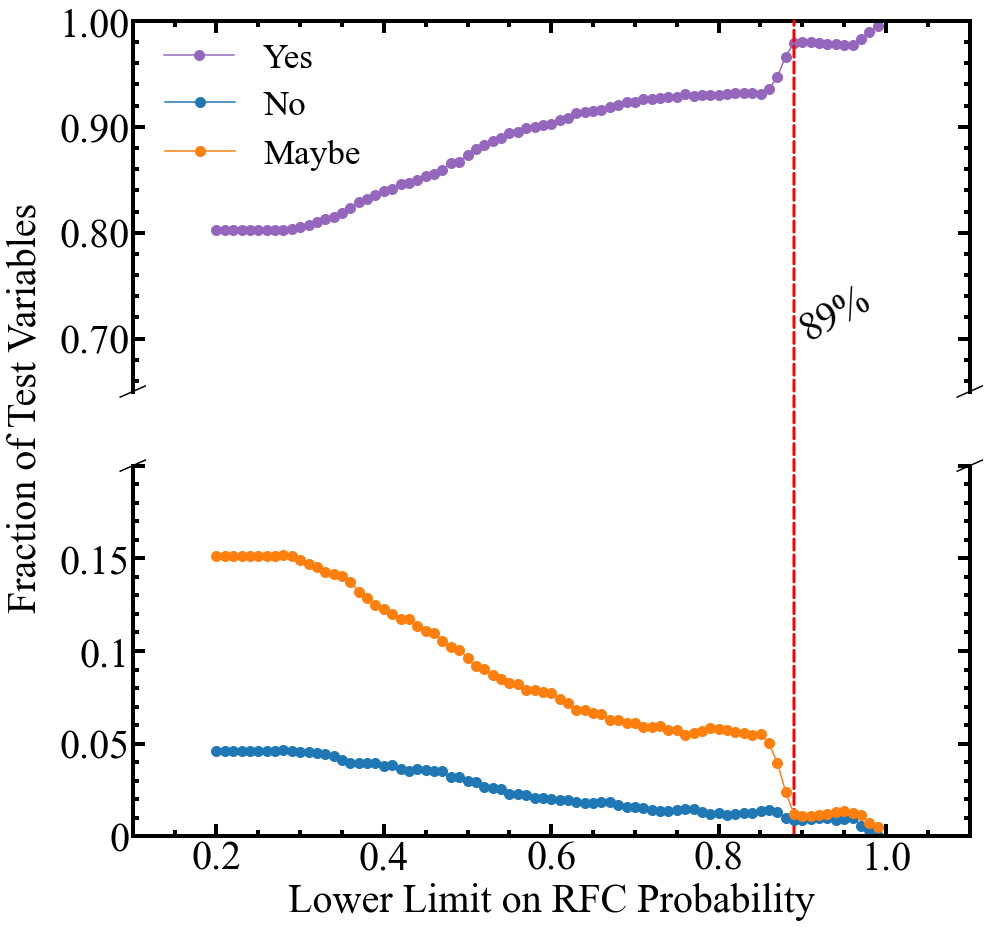}
    \caption{Integral Yes/Maybe/No fraction of test variables in our quality control sample as a function of the RFC classification probability.}
    \label{fig:cutoff}
\end{figure}

\begin{figure}
    \centering
    \includegraphics[width = 0.48\textwidth]{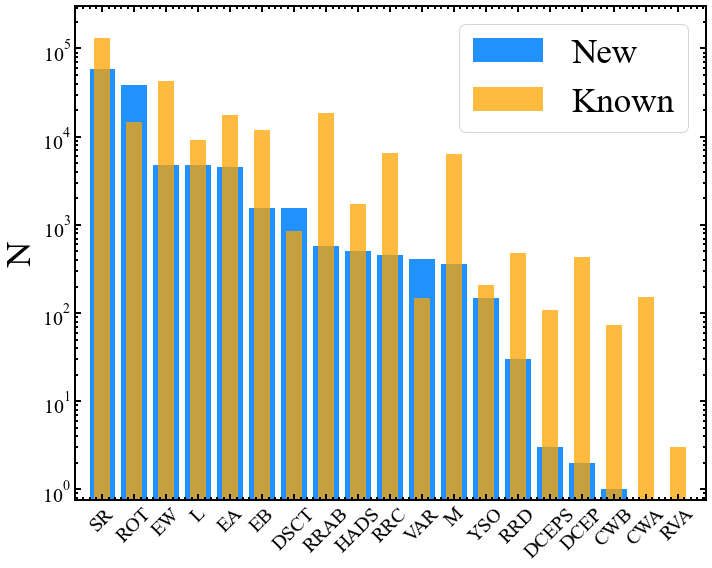}
    \caption{Number distribution of machine learning classifications after probability cuts.}
    \label{fig:categorical}
\end{figure}

\begin{figure*}
    \centering
    \includegraphics[width = 0.95\textwidth]{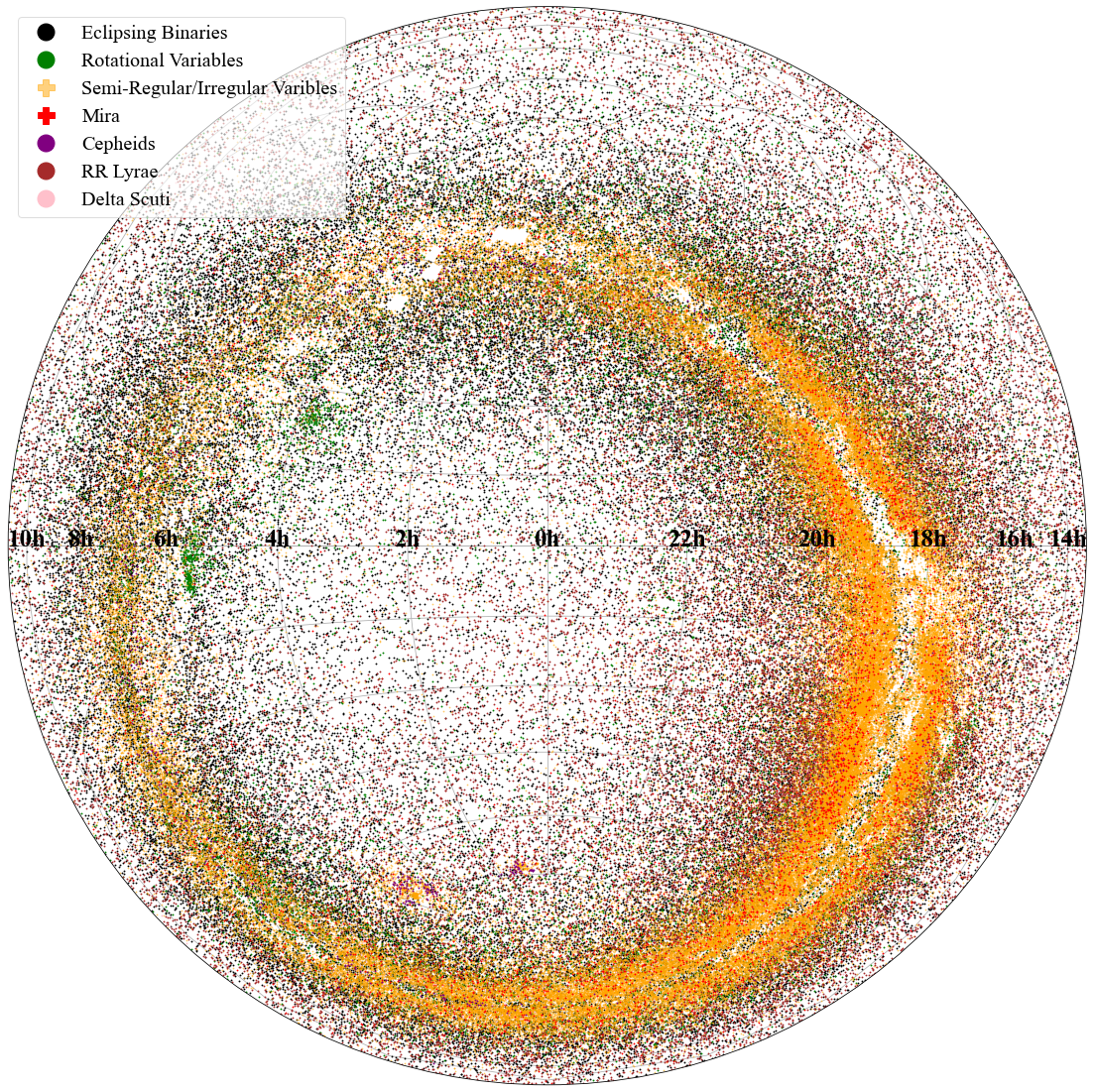}
    \caption{Equatorial distribution of the $\sim 263,000$ known variables we recovered in Equatorial coordinates.}
    \label{fig:known_sky}
\end{figure*}

\begin{figure*}
    \centering
    \includegraphics[width =  0.95\textwidth]{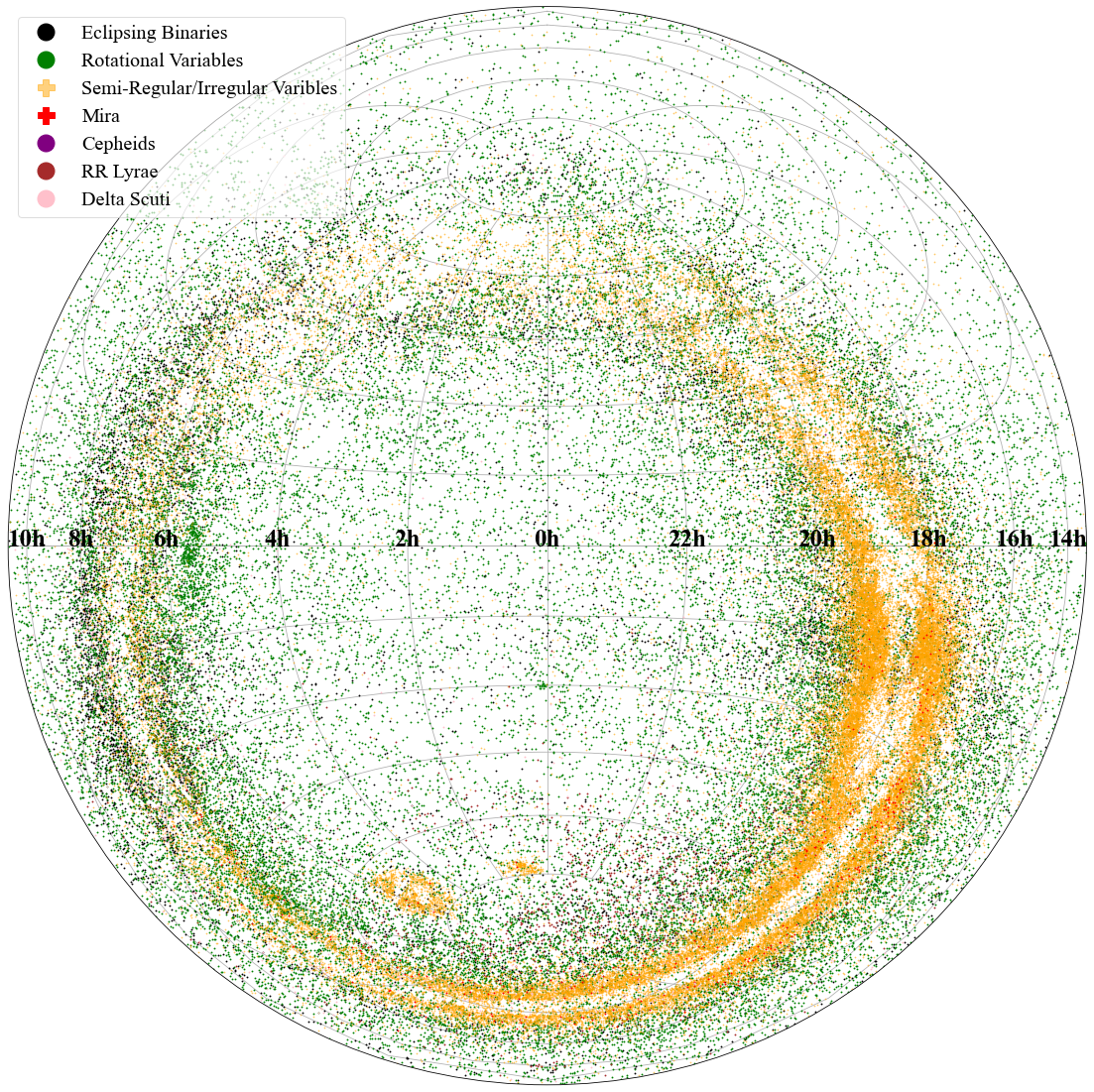}
    \caption{Equatorial distribution of the $\sim 116,000$ new variables we recovered in Equatorial coordinates.}
    \label{fig:new_sky}
\end{figure*}

We addressed this by building a private test workflow on the Zooniverse platform. ASAS-SN team members then visually inspected the light curves for 2000 randomly selected new candidates. The light curves were modelled in the same fashion as in ASAS-SN's citizen science project Citizen ASAS-SN (see \citealt{CitizenDR1}) which include versions of the following light curves: one phased using the best period, one phased using twice the best period, and the observed brightness over time. The user was asked ``Is this a variable?'' with a multiple choice response of Yes, No, and Maybe. After the test set was fully classified we found that $\sim$80\% received a ``Yes'' indicating that variability was clear while $\sim$5\% received a ``No'' and $\sim$15\% received a ``Maybe''. Figure \ref{fig:cutoff} shows how their Yes/Maybe/No classification scaled with the RFC classification probability. We decided to set the lower limit cutoff at $P \geq 0.89$, and for this limit the relative numbers are 98\% Yes, 1\% No, and 1\% Maybe. Spot checks of some of the high probability light curves voted as No and Maybe showed that these were likely user classification mistakes. Applying this cut to our full variable sample reduced the number of known and new candidates by 27\% and 71\% respectively to leave in $\sim 263,000$ known variables and $\sim 116,000$ new variables.

\section{Discussion}

The complete catalog of 378,861 variables has been added to the ASAS-SN Variable Stars Database (\url{https://asas-sn.osu.edu/variables}) along with their $g-$band light curve data and will be publicly cataloged in the AAVSO VSX catalog. The ASAS-SN catalog can be downloaded in its entirety at the ASAS-SN Variable Stars Database. 

In Table \ref{tab:1} and Figure \ref{fig:categorical} we show the number of sources for each variability type from this search. The two most common known variables found are Semi-regular (SR) and W Ursae Majoris type binaries (EW). Of the new discoveries, the most common are Semi-regular (SR) and spotted rotational variables (ROT). The number of known variable candidates was greater than the number of new candidates for all except for the ROT, DSCT, and generic VAR classes. The large number of rotating spotted stars (ROT) among our new discoveries seems to be a property of using the $g-$band instead of the $V-$band. The large rotational variability signal in the $g-$band is likely a combination of it being a bluer band and that it contains the strong Ca H and K absorption feature whose strength varies significantly with activity. We found no new W Virginis type variables (CWA), RV Tauri variables (RVA) or $\gamma$ Cassiopeiae (GCAS) type variables. There were candidates classified as GCAS variables, but they all systematically had low classification probabilities and were dropped after the cuts. The full sky distribution for several common classes of the known and new variables are shown in Figures \ref{fig:known_sky} and \ref{fig:new_sky}. The differences illustrate the large number of new rotational variables compared to the known ones. The different sky distributions are simply the differences between dwarfs and giants. The ROT variables are predominantly dwarfs and so have a relatively isotropic distribution. The SR/L variables are giants, so they trace the large scale structure of the Galaxy.

\begin{figure}
    \centering
    \includegraphics[width = 0.5\textwidth]{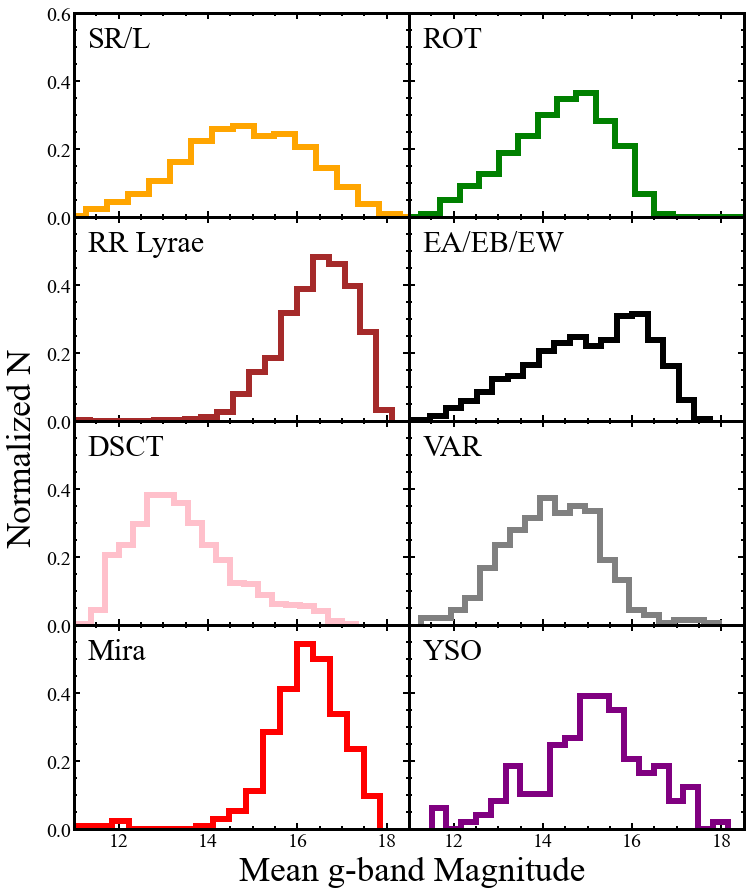}
    \caption{Distribution of the mean $g-$band magnitudes by variable class.}
    \label{fig:pulse_magdist}
\end{figure}

Figure \ref{fig:pulse_magdist} shows the distribution of the most common new variables in magnitude. Not surprisingly, most are concentrated towards fainter magnitudes since the $g-$band data are deeper than the $V-$band, with higher amplitude variables (e.g., RR Lyrae and Mira variables) being found at fainter magnitudes than lower amplitude variables (e.g., spotted stars and the generic VAR class). The two classes which do not conform to this pattern are the DSCT and SR/L variables. The DSCT variables have very short periods (hours) and their identification benefits greatly from the significantly cleaner window function for identifying short period variables in the $g-$band data. Since the ASAS-SN $g-$band light curves are the first ever to provide this level of time sampling, it is not surprising that we would identify large number of DSCT variables even at bright magnitudes. The identification of larger numbers of intermediate magnitude SR/L variables is likely a consequence of the steadily increasing total time span of the ASAS-SN light curves. We probably do not see the same thing for the Miras because their variability amplitudes are so large that they are very difficult to miss.

\begin{figure}
    \centering
    \includegraphics[width = 0.48\textwidth]{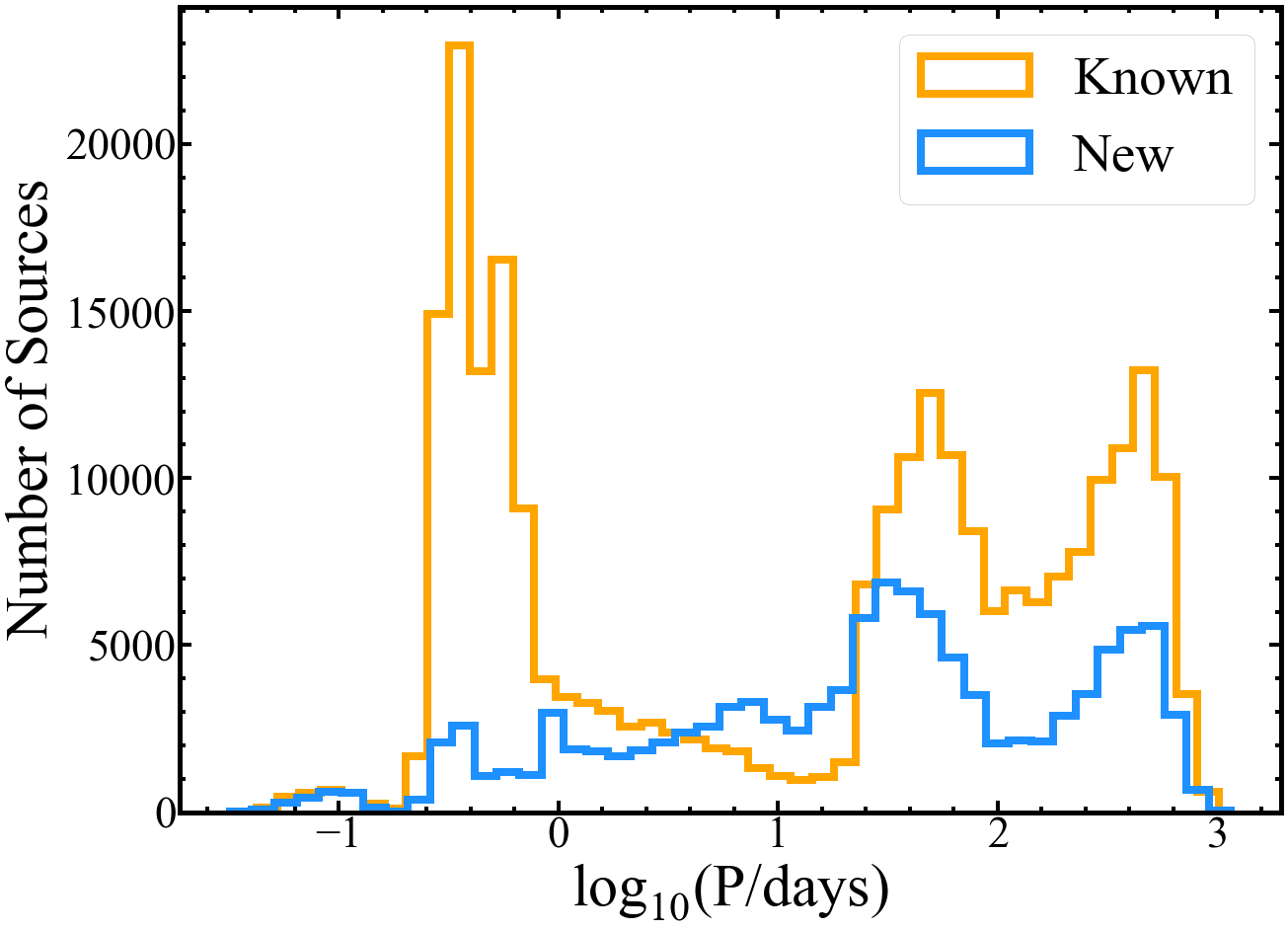}
    \caption{Number distribution of variability periods for the known and new sources.}
    \label{fig:dist_periods}
\end{figure}

In Figure \ref{fig:dist_periods} we show the period distributions for the variables. The structures at long periods are fairly similar, and are simply due to finding different relative numbers of Mira, semi-regular, and irregular variables. The lack of the large peak for periods shorter than a day is due to the low yield of RR Lyrae and EW type binaries among the new variables. These variables are comparatively easy to detect with relatively large amplitudes and easily sampled periods, so it is not surprising that the completeness of these periods surviving is high. The comparable numbers at very short periods comes from the many new DSCT variables. The larger numbers near 10 days come from the large number of rotational variables. We examined more discrete spikes close to the expected aliased periods of $\sim$1 day (diurnal alias), $\sim$14.8 day (half lunar alias), and $\sim$29.5 day (lunar alias). We spot checked a random sample of variables located near each spike and found that those with the lunar and half lunar aliased periods showed poorer signs of variability and were systematically dimmer than the sources with periods at $\sim$1 day. We suspect the dimmer sources are more prone to suffer from a low signal to noise causing the GLS periodogram to find the wrong period. To combat this, we imposed a magnitude cutoff of $g<16$ for variables with potentially aliased periods which removed the spikes at $\sim$14.8 and $\sim$29.5 days. This cut did not remove the discrete spike near $P\sim$ 1 day as most of these variables were brighter and showed clear signs of variability. We suspect that the excess of $P\sim$ 1 day period variables comes from relaxing the discrimination against diurnal aliases because of the better sampling of the $g-$band data. Citizen ASAS-SN also provided us with many examples of false positives at the diurnal aliases which went into training for the JUNK classification.

\begin{figure}
    \centering
    \includegraphics[width = 0.48\textwidth]{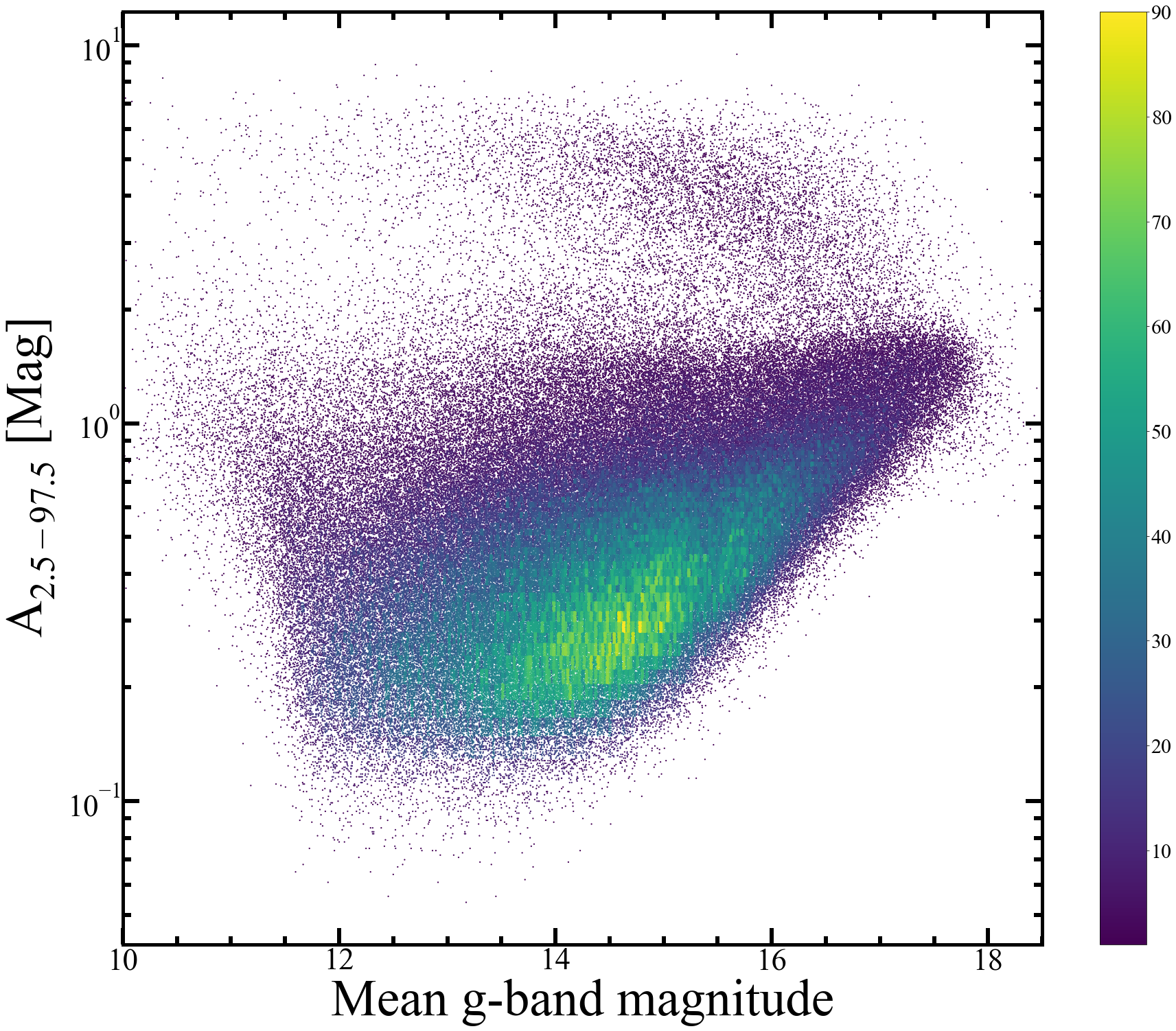}
    \caption{Distribution of variability amplitude with mean $g-$band magnitude. This includes both the new and known variables.}
    \label{fig:amp_mag}
\end{figure}

\begin{figure*}
    \centering
    \includegraphics[width =  \textwidth]{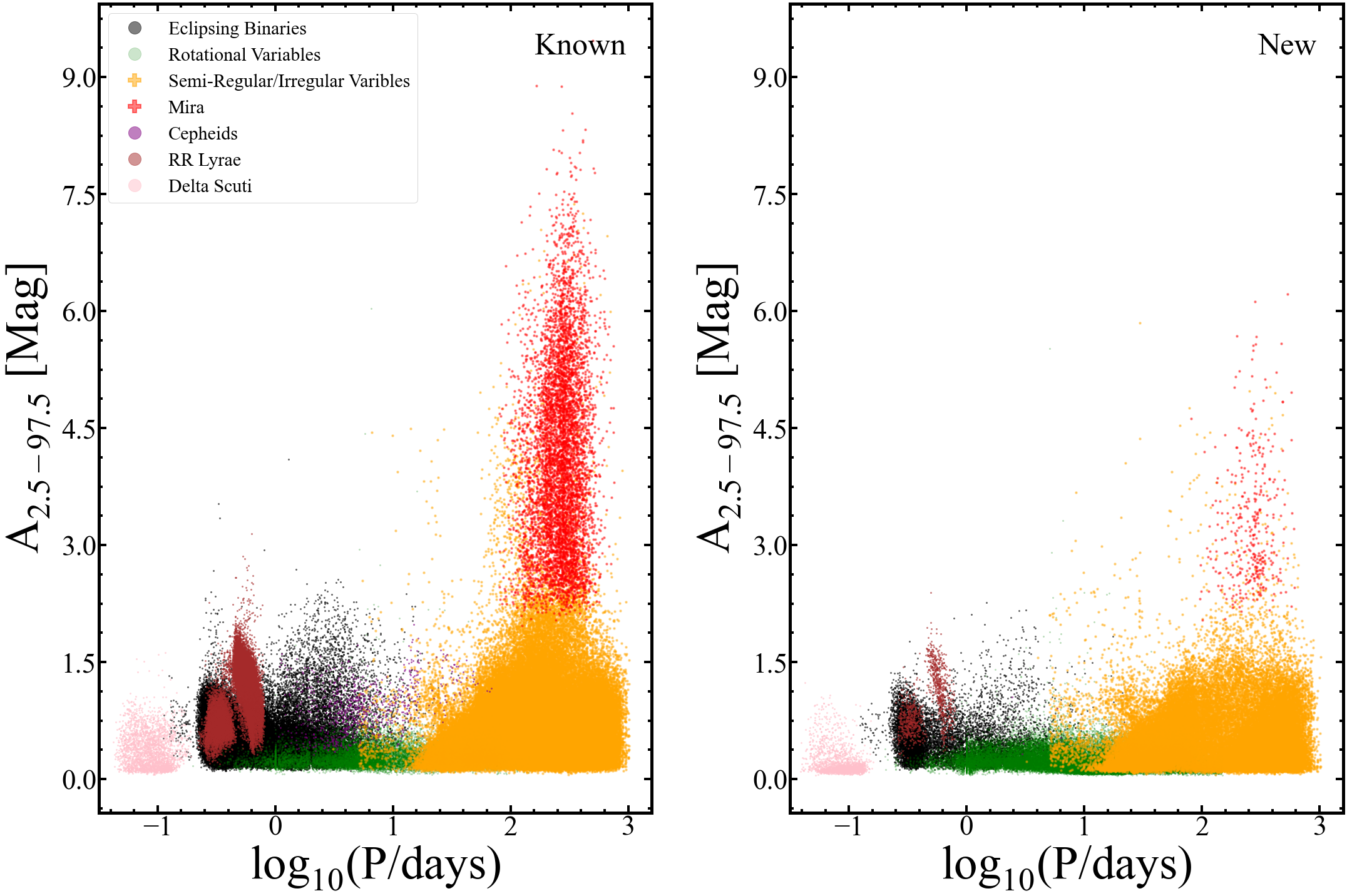}
    \caption{The $g-$band variability amplitude v.s. $\text{log}_{10}(\rm P/ \rm days)$ relationship for the known variables (Left) and new variables (Right) in our catalog using labels given by the $g-$band machine learning classifier.}
    \label{fig:amp_period}
\end{figure*}

Figure \ref{fig:amp_mag} shows the distribution of variability amplitude as a function of the mean $g-$band magnitude. The amplitude $A_{2.5-97.5}$ is the 2.5\% to the 97.5\% percentile range of the data. For bright stars, we identify variables down to amplitudes of order $\sim0.05$ mag, and then require progressively higher amplitudes beyond $g\sim16$ mag. We show the period-amplitude distribution for the known and new sources in Figure \ref{fig:amp_period}. We find relatively few bright new high amplitude pulsators, which is another indicator that searches for such variables are relatively complete.

\begin{figure*}
    \centering
    \includegraphics[width =  0.99\textwidth]{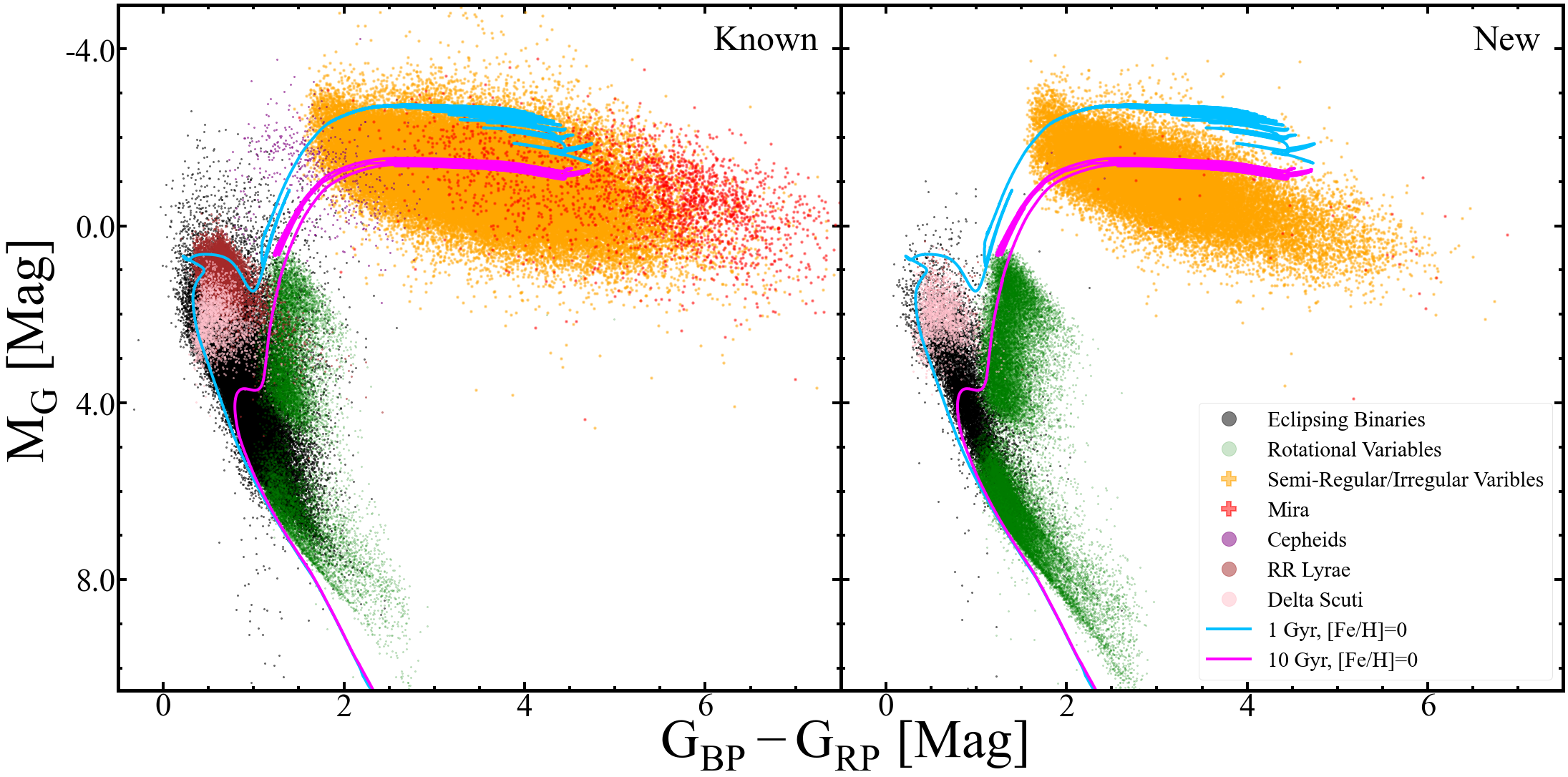}
    \caption{The \textit{Gaia} EDR3 $\rm M_G$ v.s. $\rm G_{BP}-G_{RP}$ color-magnitude diagram for the known variables (Left) and new variables (Right) in our catalog using labels given by the $g-$band machine learning classifier including 1 Gyr and 10 Gyr [Fe/H]=0 MIST isochrones.}
    \label{fig:cmd}
\end{figure*}

\begin{figure*}
    \centering
    \includegraphics[width =  \textwidth]{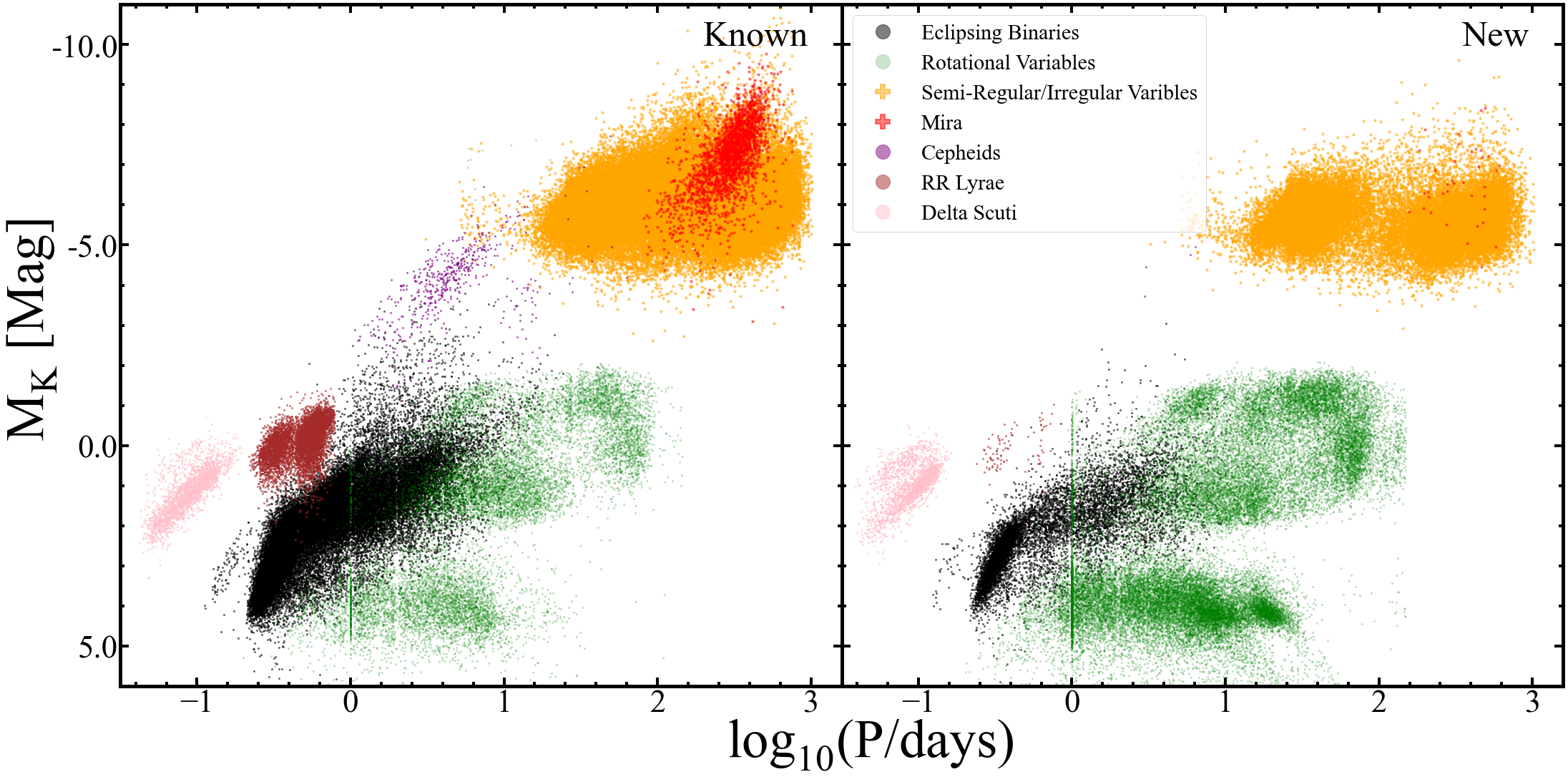}
    \caption{The $\rm M_K$ v.s. $\text{log}_{10}(\rm P/ \rm days)$ period-luminosity diagram for the known variables (Left) and new variables (Right) in our catalog using labels given by the $g-$band machine learning classifier.}
    \label{fig:plr}
\end{figure*}

Figure \ref{fig:cmd} shows the \textit{Gaia} EDR3 $\rm M_G$ v.s. $\rm G_{\rm BP}-G_{\rm RP}$ color-magnitude diagram for the known and new sources. We also overlay 1 Gyr and 10 Gyr MIST isochrones for [Fe/H]=0 to point out various evolutionary stages \citep{MIST_iso}. These tracks illustrate a lack of instability strip variables among the new variables and the abundance of spotted giants, spotted main-sequence stars, and semi-regular AGB stars. We found that a higher percentage of the new rotational variables were located on the main sequence compared to the known rotators. We also show the $\rm M_K$ v.s. $\text{log}_{10}(\rm P/ \rm days)$ period-luminosity diagram for the known and new sources in Figure \ref{fig:plr}. The two populations of semi-regular/irregular variables are clearly shown in the new variables and have been extensively observed by the OGLE survey (see \citealt{2007AcA....57..201S}). As outlined in Paper VI, the period-luminosity diagram displays two distinct populations of DSCT stars: the fundamental mode pulsators and the more luminous overtone pulsators. We found that our new $g-$band observations were able to more efficiently identify overtone DSCTs. We also found that with our larger sample of rotational variables, certain regions of the period-luminosity diagram were preferentially occupied. A more thorough analysis of the rotators will be left for a future work. The behavior of the different broad variable types and where they occupy agrees with the distribution found in the ASAS-SN $V$-band catalog \citep{Jayasinghe2019a}.

\begin{figure*}
    \centering
    \includegraphics[width = \textwidth]{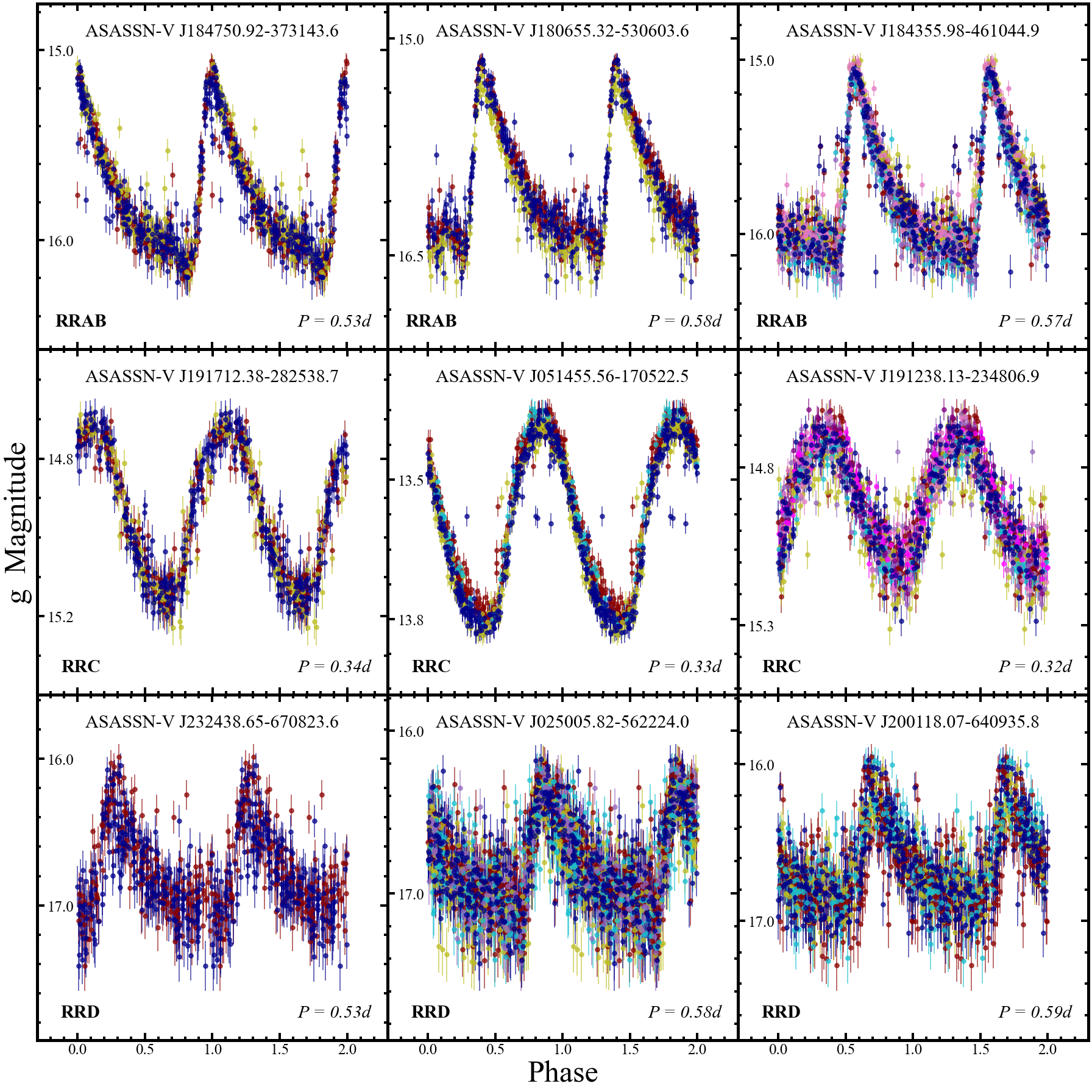}
    \caption{Phased light curves for examples of the newly discovered RR Lyrae variables. The light curves are scaled by their minimum and maximum $g-$band magnitudes. Different colored points correspond to data from the different ASAS-SN cameras. The different variability types are defined in Table \ref{tab:1}.}
    \label{fig:RR}
\end{figure*}

\begin{figure*}
    \centering
    \includegraphics[width = \textwidth]{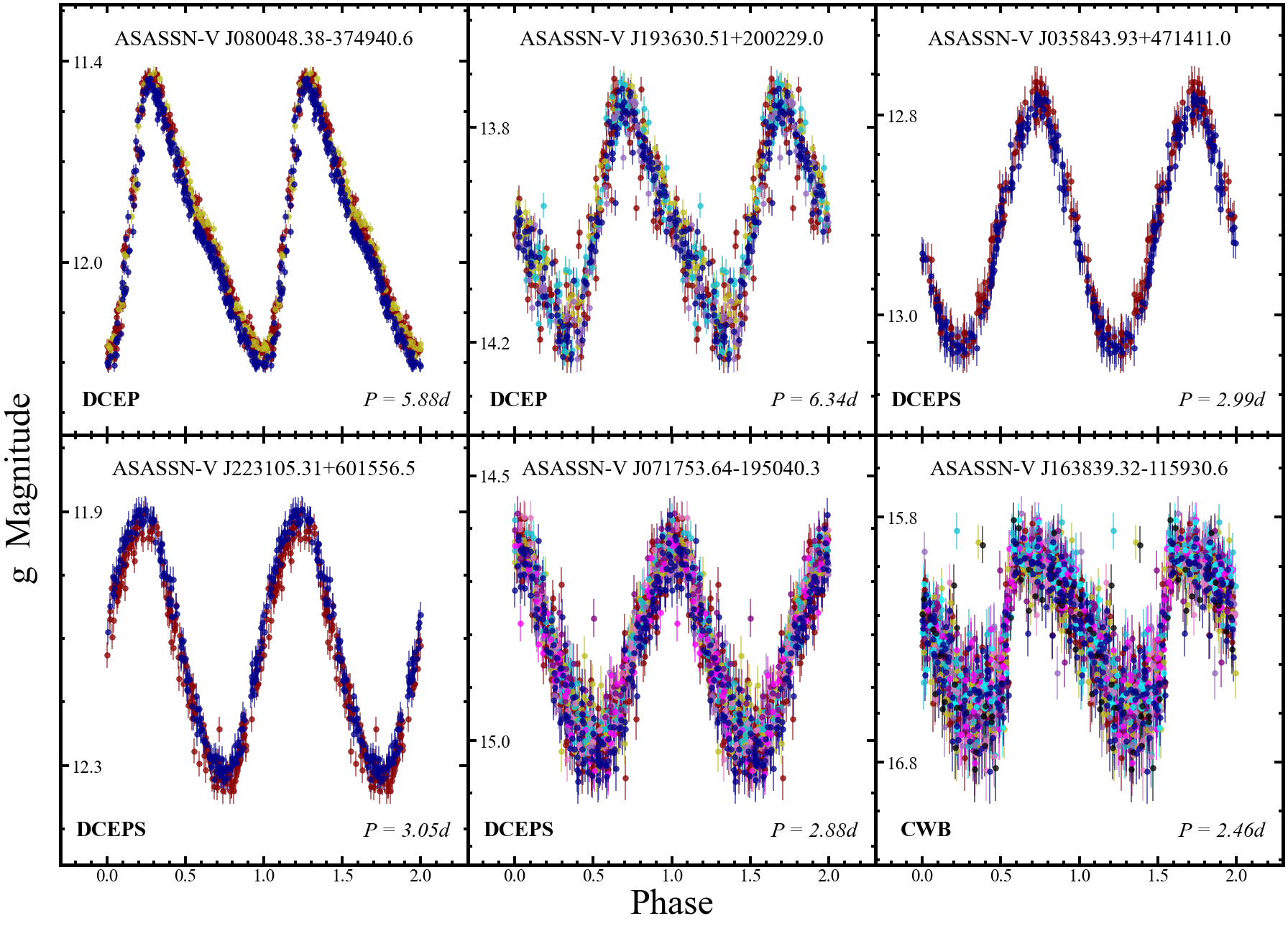}
    \caption{Light curves for the newly discovered Cepheid variables. The format is the same as for Fig \ref{fig:RR}.}
    \label{fig:Cep}
\end{figure*}

\begin{figure*}
    \centering
    \includegraphics[width = \textwidth]{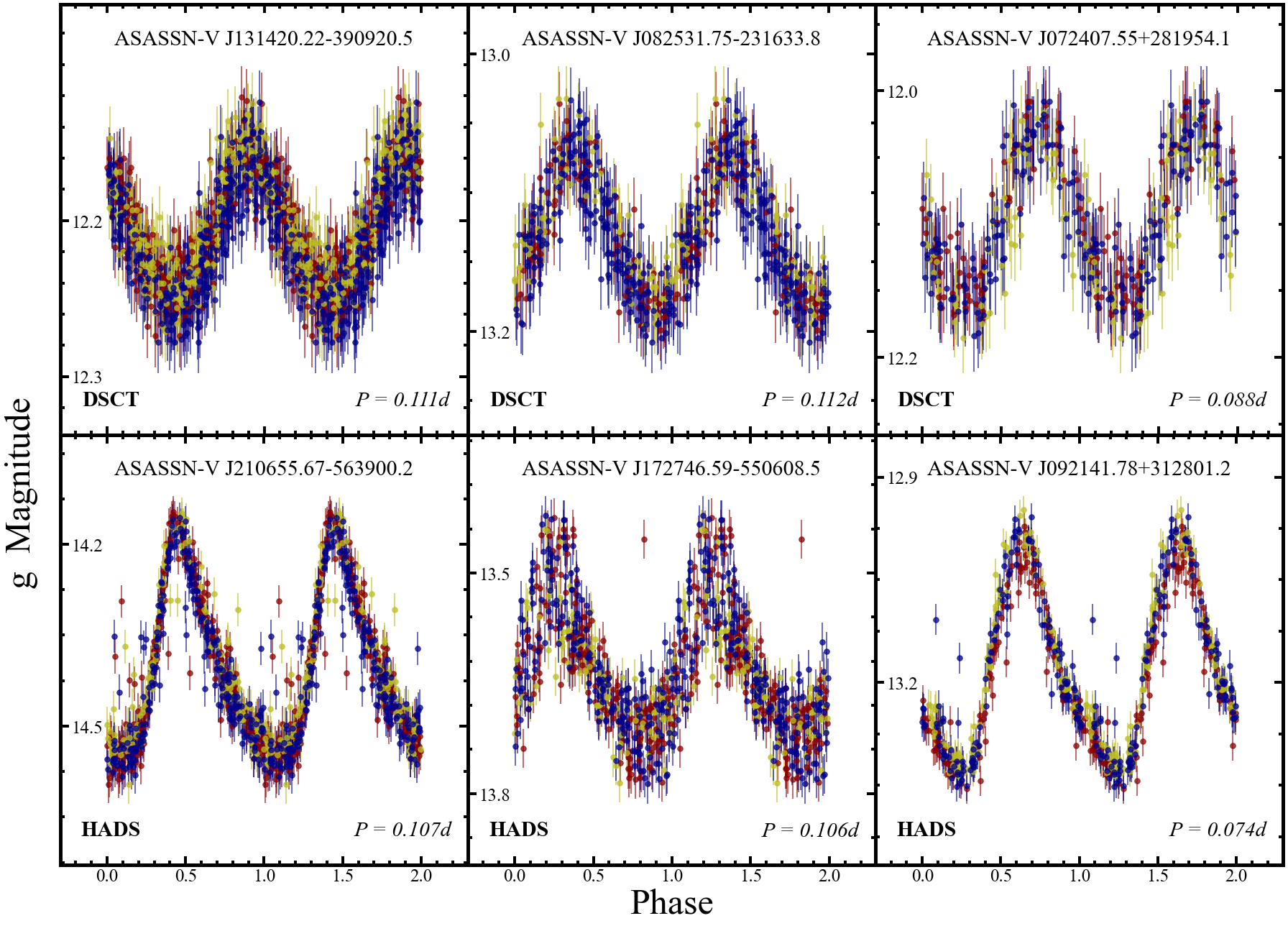}
    \caption{Light curves for examples of the newly discovered $\delta$ Scuti variables. The format is the same as for Fig \ref{fig:RR}.}
    \label{fig:DSCT}
\end{figure*}

\begin{figure*}
    \centering
    \includegraphics[width = \textwidth]{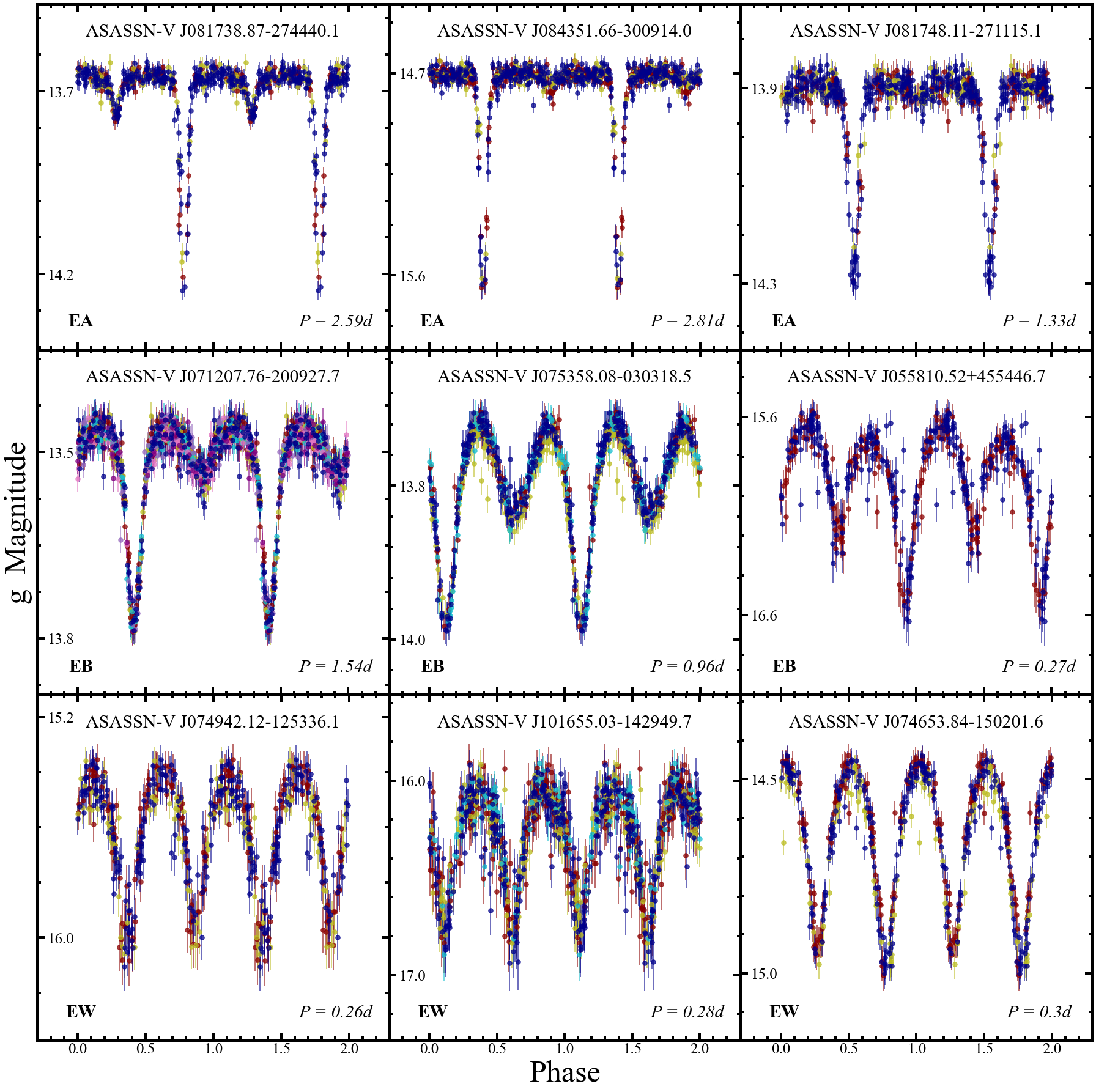}
    \caption{Light curves for examples of the newly discovered eclipsing binary variables. The format is the same as for Fig \ref{fig:RR}.}
    \label{fig:ECL}
\end{figure*}

\begin{figure*}
    \centering
    \includegraphics[width = \textwidth]{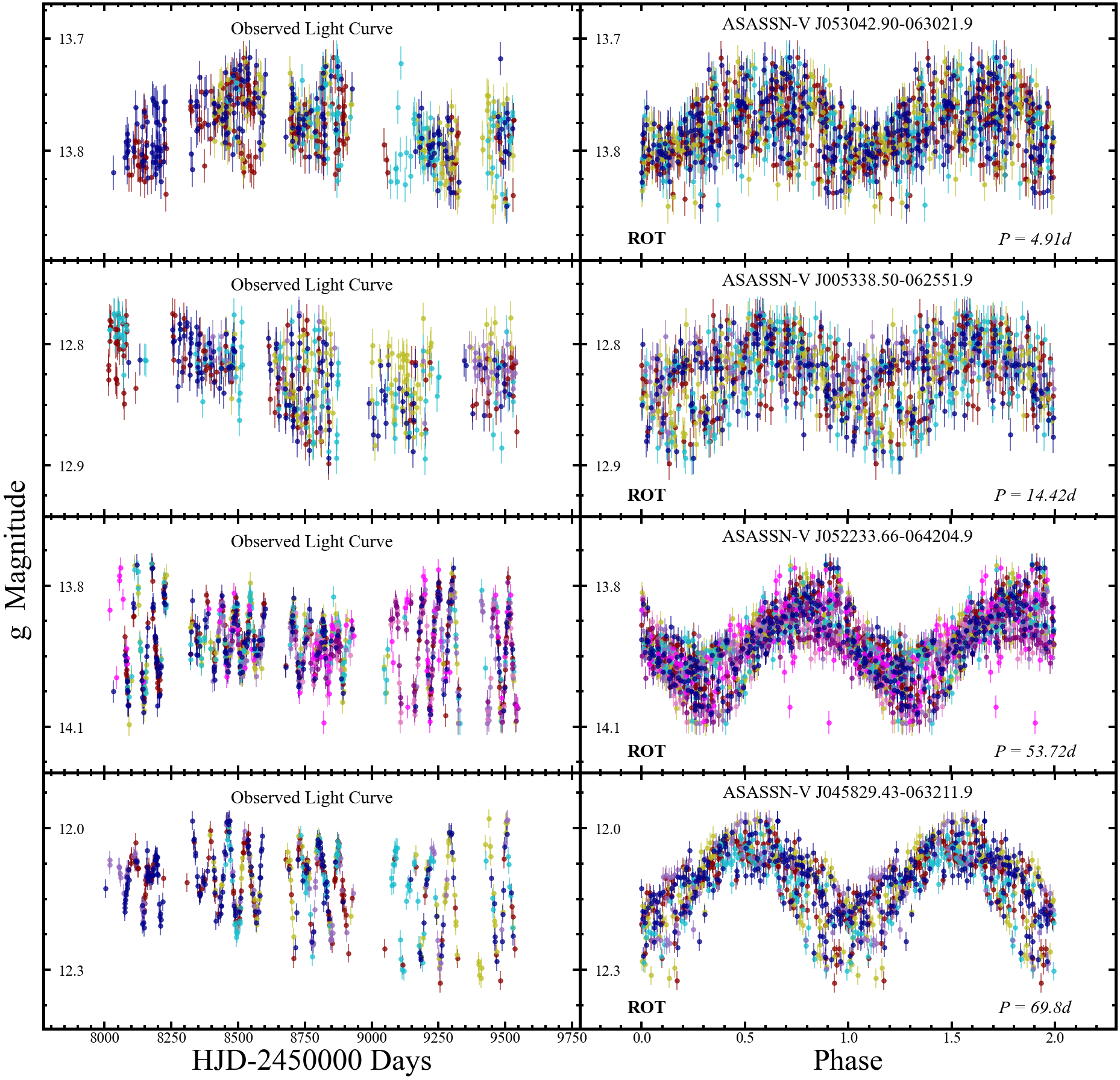}
    \caption{Light curves for examples of the newly discovered highly spotted rotational variables. The format is the same as for Fig \ref{fig:RR}.}
    \label{fig:ROT}
\end{figure*}

\begin{figure*}
    \centering
    \includegraphics[width = \textwidth]{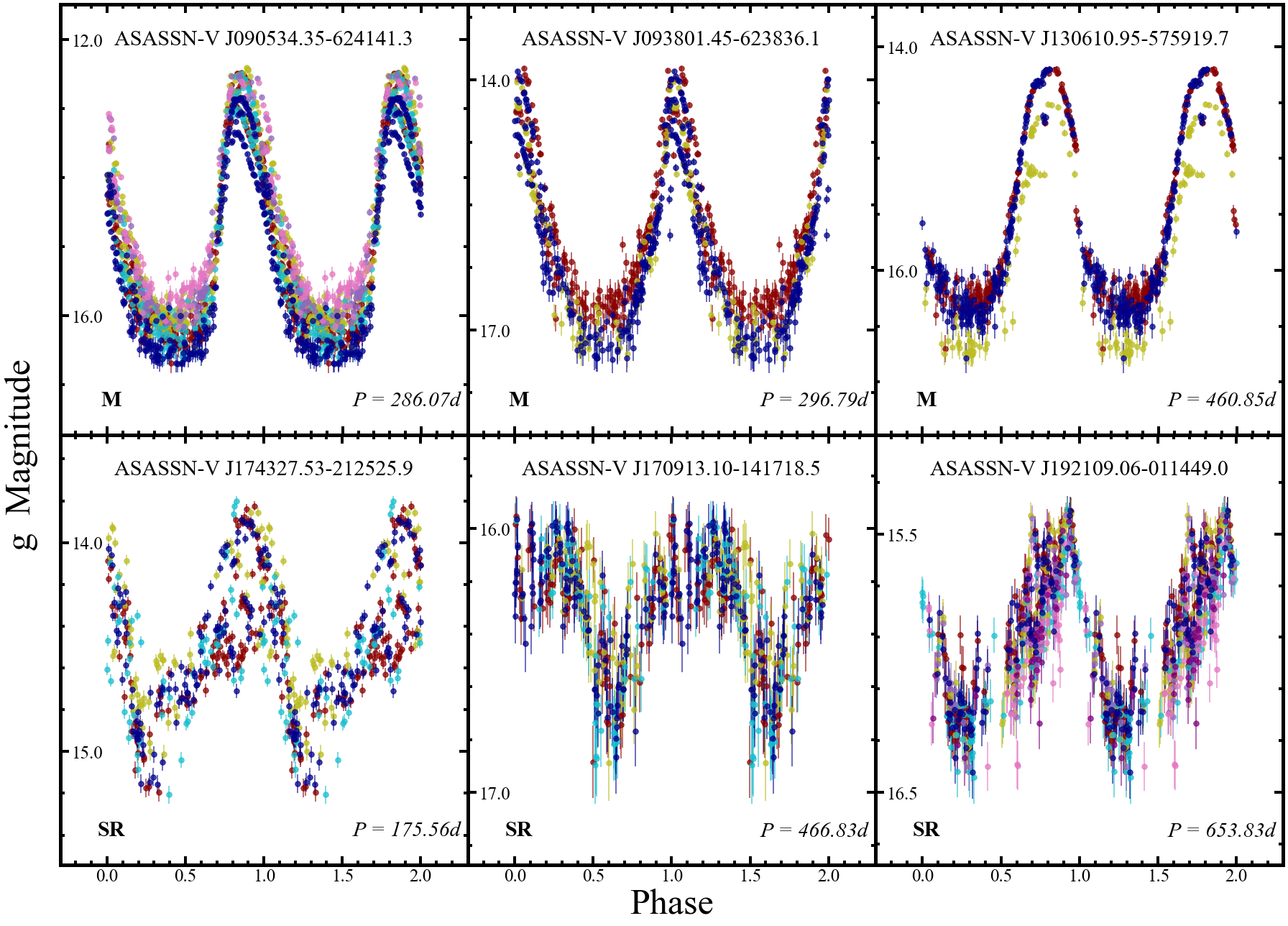}
    \caption{Light curves for examples of the newly discovered long-period variables. The format is the same as for Fig \ref{fig:RR}.}
    \label{fig:LPV}
\end{figure*}

\begin{figure*}
    \centering
    \includegraphics[width = \textwidth]{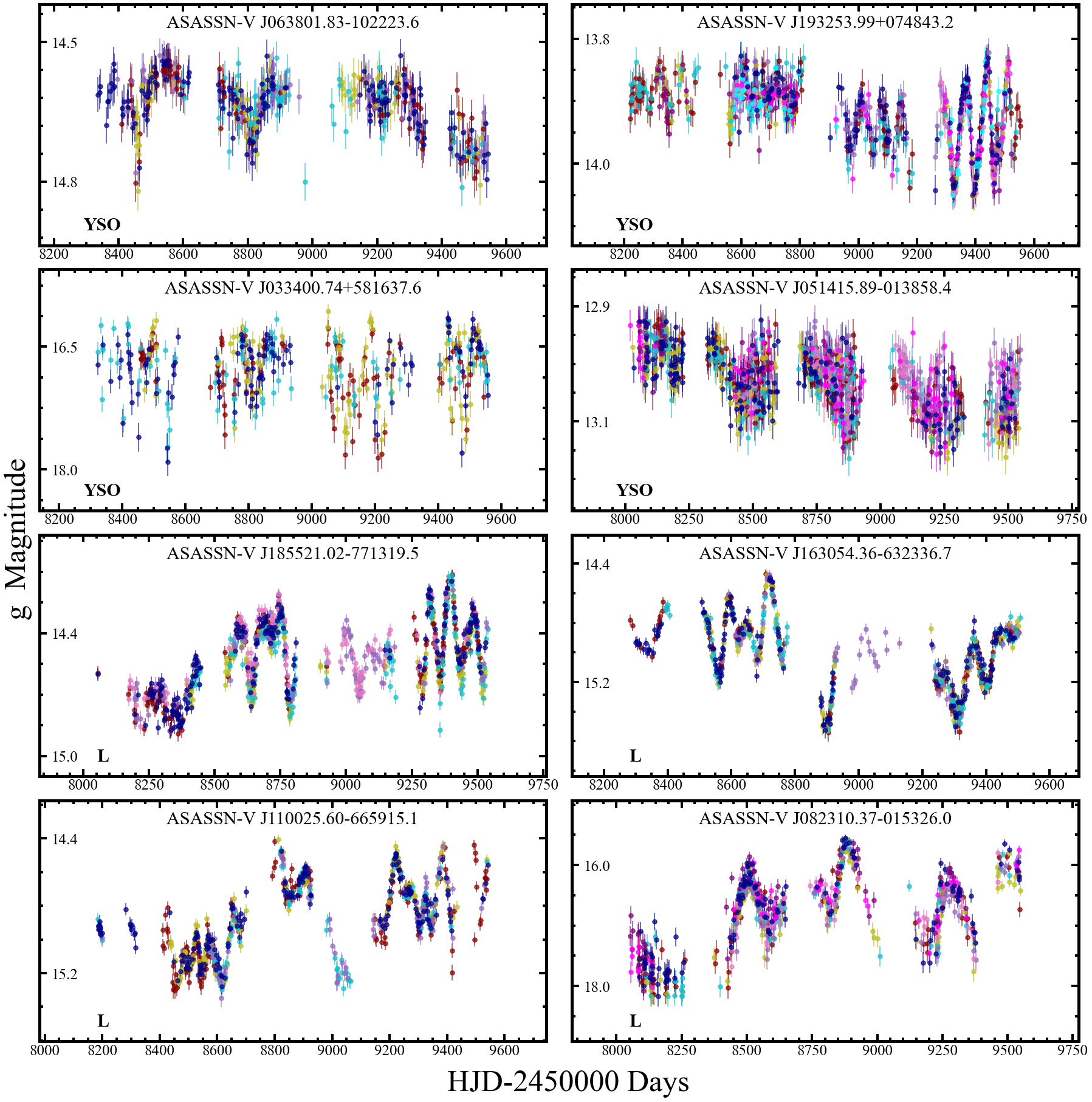}
    \caption{Light curves for examples of the newly discovered irregular variables. The format is the same as for Fig \ref{fig:RR}.}
    \label{fig:Irr}
\end{figure*}

\begin{figure*}
    \centering
    \includegraphics[width = \textwidth]{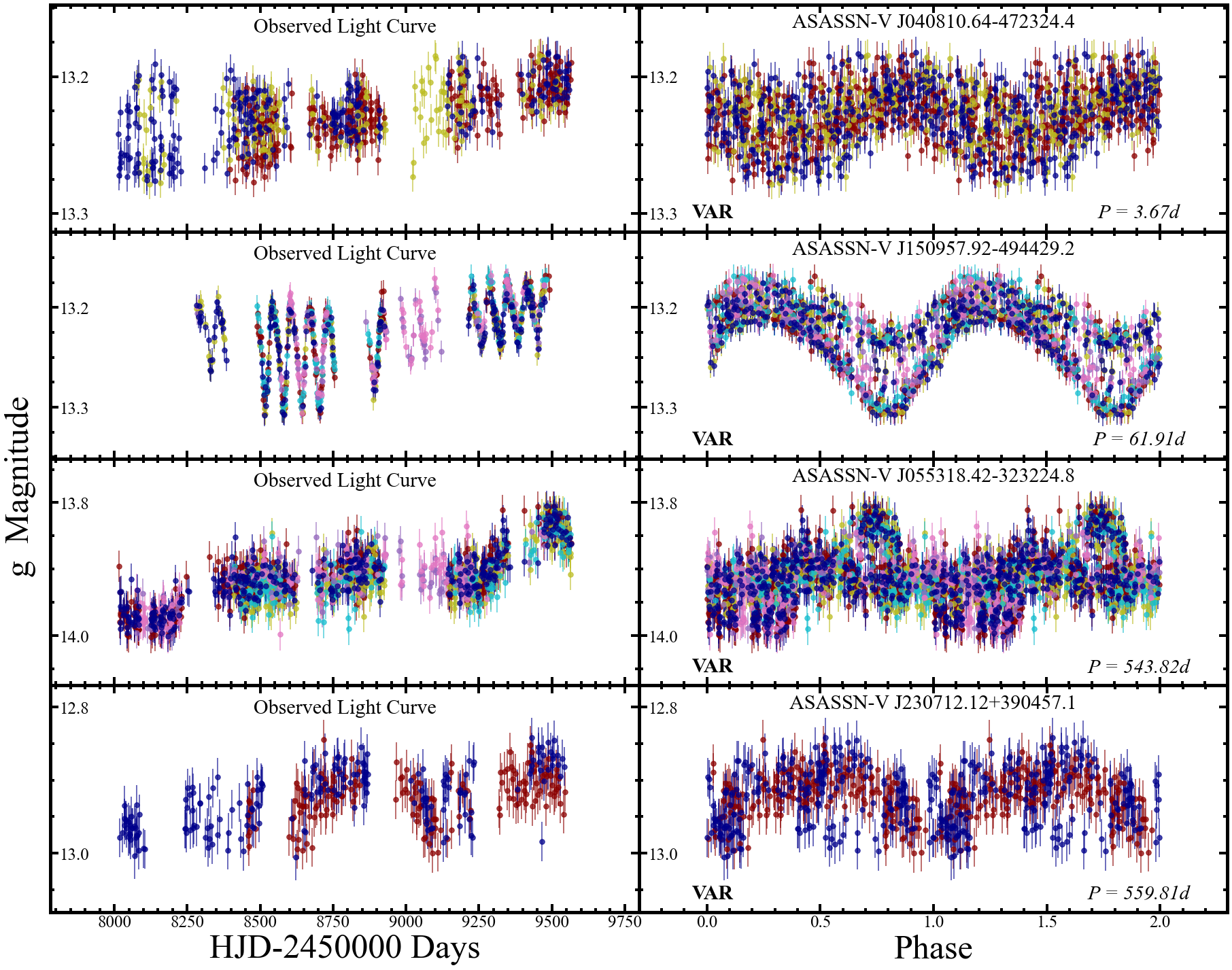}
    \caption{Light curves for examples of the newly discovered generic VAR class variables. The format is the same as for Fig \ref{fig:RR}.}
    \label{fig:VAR}
\end{figure*}

In Figure \ref{fig:RR}, we show a selection of light curves for new RR Lyrae variables. In Figure \ref{fig:Cep}, we show light curves for the new Cepheid variables. This variable class made up the smallest fraction of the new discoveries, with only 6 new sources. Our observations did not yield any new RV Tauri variables (subtype A) or W Virginis variables with periods greater than 8 days. In Figure \ref{fig:DSCT}, we show a selection of the new $\delta$ Scuti type variables. We show light curves for the new eclipsing binaries in Figure \ref{fig:ECL}, which were collectively the 3rd most common new variables. In Figure \ref{fig:ROT}, we show the phased and observed light curves for the rotational variable (ROT) class. These stars were the 2nd most common type we found and many of their observed light curves display amplitude modulation due to changes of the stars' spots with time. In Figures \ref{fig:LPV} and \ref{fig:Irr}, we show the light curves for the new long period and irregular variables. Lastly, we show light curves for new variables that were given the generic VAR class label in Figure \ref{fig:VAR}. These variables often display a clear signature of variability but the classifier was not able to assign a more specific classification.

\section{Conclusion}
In this paper, we provide an initial catalog of variable stars using the ASAS-SN $g-$band light curves. The complete list of the crossmatched variables and the ASAS-SN discoveries along with their $g-$band light curves are provided online at the ASAS-SN Variable Stars Database (\url{https://asas-sn.osu.edu/variables}). From an input catalog of $\sim$54.8 million stars, we identified $\sim$1.48 million variable candidates based on information from \textit{Gaia} EDR3, 2MASS and AllWISE. We then analyzed the light curves of these sources and found 378,861 variables, of which 262,834 are known variables and 116,027 are new discoveries. We generally recovered more known variables of each type, with the exception of the ROT, DSCT, and generic VAR classes. The most common new variables were the semi-regular and rotational variables. We found an excess of new $\sim1$ day period variables because the higher cadence and longitude spread of the $g-$band ASAS-SN configuration gives better control of the diurnal aliasing and so allows searches at these periods with fewer false positives. We also find that rotational modulations are stronger in the $g-$band, leading to many new rotational variables.

We plan to incorporate these variables, including the lower probability candidates, into our Citizen Science initiative to help refine our classifications and improve our machine learning techniques. The citizen scientists outperformed our present machine learning classifier in identifying spurious variables \citep{CitizenDR1}. Further building the JUNK training set should lead to steady improvements in the machine learning classifier. The citizen scientists also excelled at identifying unusual or extreme variable candidates. Looking forward to the \textit{Gaia} DR3 catalog, many of the variables in this catalog will be bright enough to have radial velocity measurements. \textit{Gaia}'s on-board radial velocity spectrometer (RVS) can collect radial velocities for stars brighter than $G_{\rm RVS}= 14$ mag \citep{DR3_RVS}. With this limiting magnitude, we expect many of the new discoveries in our catalog to eventually have RV measurements from \textit{Gaia}. 

\section*{Acknowledgements}

We thank the Las Cumbres Observatory and their staff for its continuing support of the ASAS-SN project. We also thank the Ohio State University College of Arts and Sciences Technology Services for helping us set up and maintain the ASAS-SN variable stars database.

ASAS-SN is funded in part by the Gordon and Betty Moore Foundation through grants GBMF5490 and GBMF10501 to the Ohio State University, and also funded in part by the Alfred P. Sloan Foundation grant G-2021-14192. TJ, KZS and CSK are supported by NSF grants AST-1814440 and AST-1908570. B.J.S. is supported by NASA grant 80NSSC19K1717 and NSF grants AST-1920392 and AST-1911074. Support for T.W.-S.H. was provided by NASA through the NASA Hubble Fellowship grant HST-HF2-51458.001-A awarded by the Space Telescope Science Institute (STScI), which is operated by the Association of Universities for Research in Astronomy, Inc., for NASA, under contract NAS5-26555. S.D. acknowledges Project 12133005 supported by National Natural Science Foundation of China (NSFC)."

Development of ASAS-SN has been supported by NSF grant AST-0908816, the Mt. Cuba Astronomical Foundation, the Center for Cosmology and AstroParticle Physics at the Ohio State University, the Chinese Academy of Sciences South America Center for Astronomy (CAS- SACA), the Villum Foundation, and George Skestos. TAT is supported in part by Scialog Scholar grant 24216 from the Research Corporation. Support for JLP is provided in part by FONDECYT through the grant 1151445 and by the Ministry of Economy, Development, and Tourism’s Millennium Science Initiative through grant IC120009, awarded to The Millennium Institute of Astrophysics, MAS.  
 
This work has made use of data from the European Space Agency (ESA) mission \textit{Gaia} (\url{https://www.cosmos.esa.int/gaia}), processed by the \textit{Gaia} Data Processing and Analysis Consortium (DPAC, \url{https://www.cosmos.esa.int/web/gaia/dpac/consortium}). Funding for the DPAC has been provided by national institutions, in particular the institutions participating in the \textit{Gaia} Multilateral Agreement.

This research has made use of the VizieR catalogue access tool, CDS, Strasbourg, France. The original description of the VizieR service was published in A$\&$AS 143, 23. 

This research made use of Astropy, a community-developed core Python package for Astronomy (Astropy Collaboration, 2013).

\section*{Data Availability}
The variables are publicly cataloged with the AAVSO and the ASAS-SN light curves can be obtained using the ASAS-SN Sky Patrol (\url{https://asas-sn.osu.edu}). The catalog of variables and the associated light curves are available on the ASAS-SN variable stars database (\url{https://asas-sn.osu.edu/variables}). The external photometric data underlying this article were accessed from sources in the public domain: \textit{Gaia} (\url{https://www.cosmos.esa.int/gaia}), 2MASS (\url{https://old.ipac.caltech.edu/2mass/overview/access.html}), AllWISE (\url{http://wise2.ipac.caltech.edu/docs/release/allwise/}) and GALEX (\url{https://archive.stsci.edu/missions-and-data/galex-1/}). 

\bibliographystyle{mnras}
\bibliography{mnras_template} 

\begin{thebibliography}{}
\makeatletter
\relax
\def\mn@urlcharsother{\let\do\@makeother \do\$\do\&\do\#\do\^\do\_\do\%\do\~}
\def\mn@doi{\begingroup\mn@urlcharsother \@ifnextchar [ {\mn@doi@}
  {\mn@doi@[]}}
\def\mn@doi@[#1]#2{\def\@tempa{#1}\ifx\@tempa\@empty \href
  {http://dx.doi.org/#2} {doi:#2}\else \href {http://dx.doi.org/#2} {#1}\fi
  \endgroup}
\def\mn@eprint#1#2{\mn@eprint@#1:#2::\@nil}
\def\mn@eprint@arXiv#1{\href {http://arxiv.org/abs/#1} {{\tt arXiv:#1}}}
\def\mn@eprint@dblp#1{\href {http://dblp.uni-trier.de/rec/bibtex/#1.xml}
  {dblp:#1}}
\def\mn@eprint@#1:#2:#3:#4\@nil{\def\@tempa {#1}\def\@tempb {#2}\def\@tempc
  {#3}\ifx \@tempc \@empty \let \@tempc \@tempb \let \@tempb \@tempa \fi \ifx
  \@tempb \@empty \def\@tempb {arXiv}\fi \@ifundefined
  {mn@eprint@\@tempb}{\@tempb:\@tempc}{\expandafter \expandafter \csname
  mn@eprint@\@tempb\endcsname \expandafter{\@tempc}}}

\bibitem[\protect\citeauthoryear{{Alard}}{{Alard}}{2000}]{2000A&AS..144..363A}
{Alard} C.,  2000, \mn@doi [\aaps] {10.1051/aas:2000214}, \href
  {https://ui.adsabs.harvard.edu/abs/2000A&AS..144..363A} {144, 363}

\bibitem[\protect\citeauthoryear{{Alard} \& {Lupton}}{{Alard} \&
  {Lupton}}{1998}]{1998ApJ...503..325A}
{Alard} C.,  {Lupton} R.~H.,  1998, \mn@doi [\apj] {10.1086/305984}, \href
  {https://ui.adsabs.harvard.edu/abs/1998ApJ...503..325A} {503, 325}

\bibitem[\protect\citeauthoryear{Alcock et~al.,}{Alcock
  et~al.}{2000}]{Alcock_2000}
Alcock C.,  et~al., 2000, \mn@doi [The Astrophysical Journal] {10.1086/309512},
  542, 281–307

\bibitem[\protect\citeauthoryear{{Andrew}, {Swihart}  \& {Strader}}{{Andrew}
  et~al.}{2021}]{Andrew2021}
{Andrew} S.,  {Swihart} S.~J.,   {Strader} J.,  2021, \mn@doi [\apj]
  {10.3847/1538-4357/abd257}, \href
  {https://ui.adsabs.harvard.edu/abs/2021ApJ...908..180A} {908, 180}

\bibitem[\protect\citeauthoryear{{Bailer-Jones}, {Rybizki}, {Fouesneau},
  {Demleitner}  \& {Andrae}}{{Bailer-Jones} et~al.}{2021}]{Bailer-Jones2021}
{Bailer-Jones} C.~A.~L.,  {Rybizki} J.,  {Fouesneau} M.,  {Demleitner} M.,
  {Andrae} R.,  2021, \mn@doi [\aj] {10.3847/1538-3881/abd806}, \href
  {https://ui.adsabs.harvard.edu/abs/2021AJ....161..147B} {161, 147}

\bibitem[\protect\citeauthoryear{Bellm}{Bellm}{2014}]{bellm2014zwicky}
Bellm E.~C.,  2014, The Zwicky Transient Facility (\mn@eprint {arXiv}
  {1410.8185})

\bibitem[\protect\citeauthoryear{{Bredall} et~al.,}{{Bredall}
  et~al.}{2020}]{2020MNRAS.496.3257B}
{Bredall} J.~W.,  et~al., 2020, \mn@doi [\mnras] {10.1093/mnras/staa1588},
  \href {https://ui.adsabs.harvard.edu/abs/2020MNRAS.496.3257B} {496, 3257}

\bibitem[\protect\citeauthoryear{Brown et~al.,}{Brown
  et~al.}{2013}]{Brown_2013}
Brown T.~M.,  et~al., 2013, \mn@doi [Publications of the Astronomical Society
  of the Pacific] {10.1086/673168}, 125, 1031–1055

\bibitem[\protect\citeauthoryear{Brown et~al.,}{Brown et~al.}{2018}]{2018}
Brown A. G.~A.,  et~al., 2018, \mn@doi [Astronomy & Astrophysics]
  {10.1051/0004-6361/201833051}, 616, A1

\bibitem[\protect\citeauthoryear{{Chen}, {Wang}, {Deng}, {de Grijs}  \&
  {Yang}}{{Chen} et~al.}{2018}]{Chen2018}
{Chen} X.,  {Wang} S.,  {Deng} L.,  {de Grijs} R.,   {Yang} M.,  2018, \mn@doi
  [\apjs] {10.3847/1538-4365/aad32b}, \href
  {https://ui.adsabs.harvard.edu/abs/2018ApJS..237...28C} {237, 28}

\bibitem[\protect\citeauthoryear{{Chen}, {Wang}, {Deng}, {de Grijs}, {Yang}  \&
  {Tian}}{{Chen} et~al.}{2020}]{Chen2020}
{Chen} X.,  {Wang} S.,  {Deng} L.,  {de Grijs} R.,  {Yang} M.,   {Tian} H.,
  2020, \mn@doi [\apjs] {10.3847/1538-4365/ab9cae}, \href
  {https://ui.adsabs.harvard.edu/abs/2020ApJS..249...18C} {249, 18}

\bibitem[\protect\citeauthoryear{{Christy} et~al.,}{{Christy}
  et~al.}{2021}]{CitizenRN}
{Christy} C.~T.,  et~al., 2021, \mn@doi [Research Notes of the American
  Astronomical Society] {10.3847/2515-5172/abe8de}, \href
  {https://ui.adsabs.harvard.edu/abs/2021RNAAS...5...38C} {5, 38}

\bibitem[\protect\citeauthoryear{{Christy} et~al.,}{{Christy}
  et~al.}{2022}]{CitizenDR1}
{Christy} C.~T.,  et~al., 2022, \mn@doi [\pasp] {10.1088/1538-3873/ac44f0},
  \href {https://ui.adsabs.harvard.edu/abs/2022PASP..134b4201C} {134, 024201}

\bibitem[\protect\citeauthoryear{Derue et~al.,}{Derue
  et~al.}{2002}]{Derue_2002}
Derue F.,  et~al., 2002, \mn@doi [Astronomy & Astrophysics]
  {10.1051/0004-6361:20020570}, 389, 149–161

\bibitem[\protect\citeauthoryear{Drake et~al.,}{Drake
  et~al.}{2009}]{Drake_2009}
Drake A.~J.,  et~al., 2009, \mn@doi [The Astrophysical Journal]
  {10.1088/0004-637x/696/1/870}, 696, 870–884

\bibitem[\protect\citeauthoryear{Feast \& Whitelock}{Feast \&
  Whitelock}{2013}]{2014IAUS..298...40F}
Feast M.,  Whitelock P.~A.,  2013, \mn@doi [Proceedings of the International
  Astronomical Union] {10.1017/s1743921313006182}, 9, 40

\bibitem[\protect\citeauthoryear{Hasanzadeh, Safari  \& Ghasemi}{Hasanzadeh
  et~al.}{2021}]{dsct}
Hasanzadeh A.,  Safari H.,   Ghasemi H.,  2021, \mn@doi [Monthly Notices of the
  Royal Astronomical Society] {10.1093/mnras/stab1411}, 505, 1476

\bibitem[\protect\citeauthoryear{{Heinze} et~al.,}{{Heinze}
  et~al.}{2018b}]{Heinze2018}
{Heinze} A.~N.,  et~al., 2018b, \mn@doi [\aj] {10.3847/1538-3881/aae47f}, \href
  {https://ui.adsabs.harvard.edu/abs/2018AJ....156..241H} {156, 241}

\bibitem[\protect\citeauthoryear{Heinze et~al.,}{Heinze
  et~al.}{2018a}]{Heinze_2018}
Heinze A.~N.,  et~al., 2018a, \mn@doi [The Astronomical Journal]
  {10.3847/1538-3881/aae47f}, 156, 241

\bibitem[\protect\citeauthoryear{Holoien et~al.,}{Holoien
  et~al.}{2016}]{Holoien_2016}
Holoien T.-S.,  et~al., 2016, \mn@doi [Monthly Notices of the Royal
  Astronomical Society] {10.1093/mnras/stw2273}, 464, 2672–2686

\bibitem[\protect\citeauthoryear{{Holtzman} et~al.,}{{Holtzman}
  et~al.}{2015}]{2015AJ....150..148H}
{Holtzman} J.~A.,  et~al., 2015, \mn@doi [\aj] {10.1088/0004-6256/150/5/148},
  \href {https://ui.adsabs.harvard.edu/abs/2015AJ....150..148H} {150, 148}

\bibitem[\protect\citeauthoryear{{Jayasinghe} et~al.,}{{Jayasinghe}
  et~al.}{2018}]{Jayasinghe2018}
{Jayasinghe} T.,  et~al., 2018, \mn@doi [\mnras] {10.1093/mnras/sty838}, \href
  {https://ui.adsabs.harvard.edu/abs/2018MNRAS.477.3145J} {477, 3145}

\bibitem[\protect\citeauthoryear{{Jayasinghe} et~al.,}{{Jayasinghe}
  et~al.}{2019a}]{Jayasinghe2019b}
{Jayasinghe} T.,  et~al., 2019a, \mn@doi [\mnras] {10.1093/mnras/stz444}, \href
  {https://ui.adsabs.harvard.edu/abs/2019MNRAS.485..961J} {485, 961}

\bibitem[\protect\citeauthoryear{{Jayasinghe} et~al.,}{{Jayasinghe}
  et~al.}{2019b}]{Jayasinghe2019a}
{Jayasinghe} T.,  et~al., 2019b, \mn@doi [\mnras] {10.1093/mnras/stz844}, \href
  {https://ui.adsabs.harvard.edu/abs/2019MNRAS.486.1907J} {486, 1907}

\bibitem[\protect\citeauthoryear{{Jayasinghe} et~al.,}{{Jayasinghe}
  et~al.}{2019c}]{Jayasinghe2019c}
{Jayasinghe} T.,  et~al., 2019c, \mn@doi [Monthly Notices of the Royal
  Astronomical Society] {10.1093/mnras/stz2711}, 491, 13

\bibitem[\protect\citeauthoryear{{Jayasinghe} et~al.,}{{Jayasinghe}
  et~al.}{2020a}]{2020arXiv200610057J}
{Jayasinghe} T.,  et~al., 2020a, arXiv e-prints, \href
  {https://ui.adsabs.harvard.edu/abs/2020arXiv200610057J} {p. arXiv:2006.10057}

\bibitem[\protect\citeauthoryear{{Jayasinghe} et~al.,}{{Jayasinghe}
  et~al.}{2020b}]{Jayasinghe2020a}
{Jayasinghe} T.,  et~al., 2020b, \mn@doi [\mnras] {10.1093/mnras/staa518},
  \href {https://ui.adsabs.harvard.edu/abs/2020MNRAS.493.4045J} {493, 4045}

\bibitem[\protect\citeauthoryear{{Jayasinghe} et~al.,}{{Jayasinghe}
  et~al.}{2020c}]{Jayasinghe2020b}
{Jayasinghe} T.,  et~al., 2020c, \mn@doi [\mnras] {10.1093/mnras/staa499},
  \href {https://ui.adsabs.harvard.edu/abs/2020MNRAS.493.4186J} {493, 4186}

\bibitem[\protect\citeauthoryear{{Jayasinghe} et~al.,}{{Jayasinghe}
  et~al.}{2021a}]{Jayasinghe2021}
{Jayasinghe} T.,  et~al., 2021a, \mn@doi [\mnras] {10.1093/mnras/stab114},
  \href {https://ui.adsabs.harvard.edu/abs/2021MNRAS.503..200J} {503, 200}

\bibitem[\protect\citeauthoryear{{Jayasinghe} et~al.,}{{Jayasinghe}
  et~al.}{2021b}]{2021MNRAS.503..200J}
{Jayasinghe} T.,  et~al., 2021b, \mn@doi [\mnras] {10.1093/mnras/stab114},
  \href {https://ui.adsabs.harvard.edu/abs/2021MNRAS.503..200J} {503, 200}

\bibitem[\protect\citeauthoryear{{Kochanek} et~al.,}{{Kochanek}
  et~al.}{2017}]{2017PASP..129j4502K}
{Kochanek} C.~S.,  et~al., 2017, \mn@doi [\pasp] {10.1088/1538-3873/aa80d9},
  \href {https://ui.adsabs.harvard.edu/abs/2017PASP..129j4502K} {129, 104502}

\bibitem[\protect\citeauthoryear{Kozlowski et~al.,}{Kozlowski
  et~al.}{2013}]{kozlowski2013supernovae}
Kozlowski S.,  et~al., 2013, Supernovae and Other Transients in the OGLE-IV
  Magellanic Bridge Data (\mn@eprint {arXiv} {1301.3909})

\bibitem[\protect\citeauthoryear{{Leavitt}}{{Leavitt}}{1908}]{1908AnHar..60...87L}
{Leavitt} H.~S.,  1908, Annals of Harvard College Observatory, \href
  {https://ui.adsabs.harvard.edu/abs/1908AnHar..60...87L} {60, 87}

\bibitem[\protect\citeauthoryear{{Lebzelter}, {Mowlavi}, {Marigo},
  {Pastorelli}, {Trabucchi}, {Wood}  \& {Lecoeur-Ta{\"\i}bi}}{{Lebzelter}
  et~al.}{2018}]{Lebzelter2018}
{Lebzelter} T.,  {Mowlavi} N.,  {Marigo} P.,  {Pastorelli} G.,  {Trabucchi} M.,
   {Wood} P.~R.,   {Lecoeur-Ta{\"\i}bi} I.,  2018, \mn@doi [\aap]
  {10.1051/0004-6361/201833615}, \href
  {https://ui.adsabs.harvard.edu/abs/2018A&A...616L..13L} {616, L13}

\bibitem[\protect\citeauthoryear{{Madore}}{{Madore}}{1982}]{Madore1982}
{Madore} B.~F.,  1982, \mn@doi [\apj] {10.1086/159659}, \href
  {https://ui.adsabs.harvard.edu/abs/1982ApJ...253..575M} {253, 575}

\bibitem[\protect\citeauthoryear{{Mateu} \& {Vivas}}{{Mateu} \&
  {Vivas}}{2018}]{2018MNRAS.479..211M}
{Mateu} C.,  {Vivas} A.~K.,  2018, \mn@doi [\mnras] {10.1093/mnras/sty1373},
  \href {https://ui.adsabs.harvard.edu/abs/2018MNRAS.479..211M} {479, 211}

\bibitem[\protect\citeauthoryear{{Pawlak} et~al.,}{{Pawlak}
  et~al.}{2019}]{2019MNRAS.487.5932P}
{Pawlak} M.,  et~al., 2019, \mn@doi [\mnras] {10.1093/mnras/stz1681}, \href
  {https://ui.adsabs.harvard.edu/abs/2019MNRAS.487.5932P} {487, 5932}

\bibitem[\protect\citeauthoryear{{Paxton} et~al.,}{{Paxton}
  et~al.}{2018}]{MIST_iso}
{Paxton} B.,  et~al., 2018, \mn@doi [\apjs] {10.3847/1538-4365/aaa5a8}, \href
  {https://ui.adsabs.harvard.edu/abs/2018ApJS..234...34P} {234, 34}

\bibitem[\protect\citeauthoryear{{Pedregosa} et~al.,}{{Pedregosa}
  et~al.}{2012}]{2012arXiv1201.0490P}
{Pedregosa} F.,  et~al., 2012, arXiv e-prints, \href
  {https://ui.adsabs.harvard.edu/abs/2012arXiv1201.0490P} {p. arXiv:1201.0490}

\bibitem[\protect\citeauthoryear{{Percy}}{{Percy}}{2007}]{Percy}
{Percy} J.~R.,  2007, {Understanding Variable Stars}.
Cambridge University Press ({CUP})

\bibitem[\protect\citeauthoryear{{Pietrukowicz} et~al.,}{{Pietrukowicz}
  et~al.}{2020}]{Pietrukowicz2020}
{Pietrukowicz} P.,  et~al., 2020, \mn@doi [\actaa] {10.32023/0001-5237/70.4.1},
  \href {https://ui.adsabs.harvard.edu/abs/2020AcA....70..241P} {70, 241}

\bibitem[\protect\citeauthoryear{Pojmanski}{Pojmanski}{2002}]{pojmanski_2002}
Pojmanski G.,  2002, The All Sky Automated Survey. Variable Stars in the 0h -
  6h Quarter of the Southern Hemisphere, \url
  {https://arxiv.org/abs/astro-ph/0210283}

\bibitem[\protect\citeauthoryear{Poleski, Soszyński, Udalski, Szymański,
  Kubiak, Pietrzynski, Wyrzykowski  \& Ulaczyk}{Poleski
  et~al.}{2012}]{poleski2012optical}
Poleski R.,  Soszyński I.,  Udalski A.,  Szymański M.~K.,  Kubiak M.,
  Pietrzynski G.,  Wyrzykowski L.,   Ulaczyk K.,  2012, The Optical
  Gravitational Lensing Experiment. The Catalog of Stellar Proper Motions
  toward the Magellanic Clouds (\mn@eprint {arXiv} {1203.2649})

\bibitem[\protect\citeauthoryear{Prusti et~al.,}{Prusti et~al.}{2016}]{2016}
Prusti T.,  et~al., 2016, \mn@doi [Astronomy & Astrophysics]
  {10.1051/0004-6361/201629272}, 595, A1

\bibitem[\protect\citeauthoryear{{Ricker} et~al.,}{{Ricker}
  et~al.}{2015}]{2015JATIS...1a4003R}
{Ricker} G.~R.,  et~al., 2015, \mn@doi [Journal of Astronomical Telescopes,
  Instruments, and Systems] {10.1117/1.JATIS.1.1.014003}, \href
  {https://ui.adsabs.harvard.edu/abs/2015JATIS...1a4003R} {1, 014003}

\bibitem[\protect\citeauthoryear{{Seabroke} et~al.,}{{Seabroke}
  et~al.}{2021}]{DR3_RVS}
{Seabroke} G.~M.,  et~al., 2021, \mn@doi [\aap] {10.1051/0004-6361/202141008},
  \href {https://ui.adsabs.harvard.edu/abs/2021A&A...653A.160S} {653, A160}

\bibitem[\protect\citeauthoryear{{Shappee} et~al.,}{{Shappee}
  et~al.}{2014}]{2014ApJ...788...48S}
{Shappee} B.~J.,  et~al., 2014, \mn@doi [\apj] {10.1088/0004-637X/788/1/48},
  \href {https://ui.adsabs.harvard.edu/abs/2014ApJ...788...48S} {788, 48}

\bibitem[\protect\citeauthoryear{{Soszy{\'n}ski} et~al.,}{{Soszy{\'n}ski}
  et~al.}{2007}]{2007AcA....57..201S}
{Soszy{\'n}ski} I.,  et~al., 2007, \actaa, \href
  {https://ui.adsabs.harvard.edu/abs/2007AcA....57..201S} {57, 201}

\bibitem[\protect\citeauthoryear{{Soszy{\'n}ski} et~al.,}{{Soszy{\'n}ski}
  et~al.}{2014}]{Soszynski2014}
{Soszy{\'n}ski} I.,  et~al., 2014, \actaa, \href
  {https://ui.adsabs.harvard.edu/abs/2014AcA....64..177S} {64, 177}

\bibitem[\protect\citeauthoryear{{Soszy{\'n}ski} et~al.,}{{Soszy{\'n}ski}
  et~al.}{2015}]{Soszynski2015}
{Soszy{\'n}ski} I.,  et~al., 2015, \actaa, \href
  {https://ui.adsabs.harvard.edu/abs/2015AcA....65..297S} {65, 297}

\bibitem[\protect\citeauthoryear{{Soszy{\'n}ski} et~al.,}{{Soszy{\'n}ski}
  et~al.}{2016}]{Soszynski2016}
{Soszy{\'n}ski} I.,  et~al., 2016, \actaa, \href
  {https://ui.adsabs.harvard.edu/abs/2016AcA....66..405S} {66, 405}

\bibitem[\protect\citeauthoryear{{Soszy{\'n}ski} et~al.,}{{Soszy{\'n}ski}
  et~al.}{2021}]{Soszynski2021}
{Soszy{\'n}ski} I.,  et~al., 2021, \mn@doi [\actaa]
  {10.32023/0001-5237/71.3.1}, \href
  {https://ui.adsabs.harvard.edu/abs/2021AcA....71..189S} {71, 189}

\bibitem[\protect\citeauthoryear{{Thompson} et~al.,}{{Thompson}
  et~al.}{2019}]{2019Sci...366..637T}
{Thompson} T.~A.,  et~al., 2019, \mn@doi [Science] {10.1126/science.aau4005},
  \href {https://ui.adsabs.harvard.edu/abs/2019Sci...366..637T} {366, 637}

\bibitem[\protect\citeauthoryear{Tonry et~al.,}{Tonry
  et~al.}{2018a}]{Tonry_2018}
Tonry J.~L.,  et~al., 2018a, \mn@doi [Publications of the Astronomical Society
  of the Pacific] {10.1088/1538-3873/aabadf}, 130, 064505

\bibitem[\protect\citeauthoryear{{Tonry} et~al.,}{{Tonry}
  et~al.}{2018b}]{2018ApJ...867..105T}
{Tonry} J.~L.,  et~al., 2018b, \mn@doi [\apj] {10.3847/1538-4357/aae386}, \href
  {https://ui.adsabs.harvard.edu/abs/2018ApJ...867..105T} {867, 105}

\bibitem[\protect\citeauthoryear{Torres, Andersen  \& Giménez}{Torres
  et~al.}{2009}]{Torres_2009}
Torres G.,  Andersen J.,   Giménez A.,  2009, \mn@doi [The Astronomy and
  Astrophysics Review] {10.1007/s00159-009-0025-1}, 18, 67–126

\bibitem[\protect\citeauthoryear{Udalski}{Udalski}{2004}]{udalski_2004}
Udalski A.,  2004, The Optical Gravitational Lensing Experiment. Real Time Data
  Analysis Systems in the OGLE-III Survey, \url
  {https://arxiv.org/abs/astro-ph/0401123}

\bibitem[\protect\citeauthoryear{{Udalski}, {Szyma{\'n}ski}  \&
  {Szyma{\'n}ski}}{{Udalski} et~al.}{2015}]{Udalski2015}
{Udalski} A.,  {Szyma{\'n}ski} M.~K.,   {Szyma{\'n}ski} G.,  2015, \actaa,
  \href {https://ui.adsabs.harvard.edu/abs/2015AcA....65....1U} {65, 1}

\bibitem[\protect\citeauthoryear{{Udalski} et~al.,}{{Udalski}
  et~al.}{2018}]{Udalski2018}
{Udalski} A.,  et~al., 2018, \mn@doi [\actaa] {10.32023/0001-5237/68.4.1},
  \href {https://ui.adsabs.harvard.edu/abs/2018AcA....68..315U} {68, 315}

\bibitem[\protect\citeauthoryear{{Watson}, {Henden}  \& {Price}}{{Watson}
  et~al.}{2006}]{2006SASS...25...47W}
{Watson} C.~L.,  {Henden} A.~A.,   {Price} A.,  2006, Society for Astronomical
  Sciences Annual Symposium, \href
  {https://ui.adsabs.harvard.edu/abs/2006SASS...25...47W} {25, 47}

\bibitem[\protect\citeauthoryear{Woźniak et~al.,}{Woźniak
  et~al.}{2004}]{Woniak_2004}
Woźniak P.~R.,  et~al., 2004, \mn@doi [The Astronomical Journal]
  {10.1086/382719}, 127, 2436–2449

\makeatother
\end{thebibliography}
\bsp	
\label{lastpage}
\end{document}